\documentclass[11pt,a4paper]{article}
\pdfoutput=1 
\usepackage{jcappub}

\usepackage{ifpdf}
\usepackage{amsfonts}
\usepackage{mathtools}
\usepackage{braket}
\usepackage{tensor}
\usepackage{bm}
\usepackage{slashed}
\usepackage{enumerate}
\usepackage{feynmp}
\usepackage{multirow}
\usepackage{xcolor}
\usepackage{tabls}
\usepackage{dcolumn}

\begin{document}


\title{Lattice study of primordial black hole formation in bumpy axion inflation}
\author[a,b,c]{Masahiro Kawasaki,}
\author[d,e]{Kai Murai,}
\author[b]{Shunsuke Tsuchida}

\affiliation[a]{Kavli IPMU (WPI), UTIAS, University of Tokyo, Kashiwa 277-8583, Japan}
\affiliation[b]{ICRR, University of Tokyo, Kashiwa 277-8582, Japan}
\affiliation[c]{Physics Division, Faculty of Science, Kanagawa University, Kanagawa 221-8686, Japan}
\affiliation[d]{RIKEN Center for Interdisciplinary Theoretical and Mathematical Sciences (iTHEMS), RIKEN, Wako 351-0198, Japan}
\affiliation[e]{Department of Physics, Tohoku University, Sendai 980-8578, Japan}

\abstract{
We study primordial black hole (PBH) formation in axion $U(1)$ inflation using lattice simulations.
In axion $U(1)$ inflation with a bumpy potential, the curvature perturbations can be enhanced in a narrow range of wavenumbers, potentially leading to PBH formation.
After confirming that our lattice simulations reproduced the known curvature power spectra for chaotic inflation and simple axion $U(1)$ inflation, we calculate the curvature power spectrum in the bumpy axion inflation model in the strong backreaction regime.
We find that large curvature perturbations are generated, which lead to PBH production with an abundance sufficient to account for dark matter.
}


\emailAdd{masahiro.kawasaki@ipmu.jp}
\emailAdd{kai.murai@riken.jp}

\begin{flushright}
    RIKEN-iTHEMS-Report-26
\end{flushright}
\maketitle

\section{Introduction}
\label{sec: intro}

Primordial black holes (PBHs) formed from large density fluctuations in the early universe~\cite{Hawking:1971ei,Carr:1974nx,Carr:1975qj} have attracted much attention in cosmology, astrophysics, and particle physics since they could account for the dark matter of the universe, the gravitational-wave events observed by the LIGO-Virgo-KAGRA collaboration~\cite{KAGRA:2021vkt}, and the seeds of supermassive black holes~\cite{Kormendy:1995er,Magorrian:1997hw,Richstone:1998ky}. 
Many inflationary models for producing PBHs~\cite{Frampton:2010sw,Kawasaki:2012kn,Clesse:2015wea,Kawasaki:2016pql,Inomata:2016rbd,Inomata:2017okj,Garcia-Bellido:2017mdw,Di:2017ndc,Inomata:2018cht,Dalianis:2018frf,Bhaumik:2019tvl,Atal:2019erb,Mishra:2019pzq,Braglia:2020fms,Fumagalli:2020adf,Braglia:2020eai,Fu:2020lob,Ragavendra:2020sop,Anguelova:2020nzl,Heydari:2021gea,Bhattacharya:2022fze,Cai:2022erk,Pi:2022zxs,Qin:2023lgo,Tada:2023pue,Cai:2023uhc,Heydari:2023rmq,Choudhury:2024one,Dimastrogiovanni:2024xvc,Wang:2024vfv} have been proposed since the early studies~\cite{Ivanov:1994pa,Yokoyama:1995ex,GarciaBellido:1996qt,Kawasaki:1997ju,Yokoyama:1998pt}.

In this paper, we focus on the axion $U(1)$ inflation model~\cite{Anber:2006xt,Anber:2009ua,Barnaby:2010vf}, in which an axion-like field, i.e., a pseudo-Nambu-Goldstone field, plays the role of an inflaton and couples to a $U(1)$ gauge field through the Chern-Simons interaction.
In this model, one of the polarization modes of the gauge field becomes tachyonically unstable, which leads to rapid production of the gauge particles due to the motion of the inflaton field.
The produced gauge particles then source inflaton fluctuations.
They also backreact on the evolution of the homogeneous part of the inflaton field and the Hubble parameter.
Both effects enhance the curvature perturbations, and consequently, a substantial amount of PBHs can be produced in the axion $U(1)$ inflation model~\cite{Linde:2012bt,Bugaev:2013fya,Garcia-Bellido:2016dkw,Domcke:2017fix,Cheng:2018yyr,Almeida:2020kaq,Unal:2023srk,Franciolini:2026cps}.
Furthermore, the enhanced curvature perturbations can source primordial gravitational waves with potentially observable amplitudes~\cite{Sorbo:2011rz,Barnaby:2011vw,Cook:2011hg,Barnaby:2011qe,Barnaby:2012xt,Cook:2013xea,Mukohyama:2014gba,Ferreira:2014zia,Namba:2015gja,Domcke:2016bkh,Garcia-Bellido:2016dkw,Obata:2016oym,Cheng:2018yyr,Ozsoy:2020ccy,Almeida:2020kaq,Unal:2023srk,Garcia-Bellido:2023ser,Corba:2024tfz,Ozsoy:2024apn,Corba:2025reo,vonEckardstein:2025oic,Teuscher:2025jhq,Barbon:2025wjl}.

Since the gauge-field production is a nonlinear phenomenon, we need lattice simulations to precisely calculate the evolution of the inflaton and gauge fields during the inflationary epoch. 
However, lattice studies of such inflationary dynamics were absent for a long time.
Recently, Refs.~\cite{Caravano:2021pgc,Caravano:2021bfn,Caravano:2022epk,Figueroa:2023oxc,Caravano:2024xsb,Sharma:2024nfu,Figueroa:2024rkr,Lizarraga:2025aiw,Jamieson:2025ngu} studied axion $U(1)$ inflation using lattice simulation and evaluated the curvature perturbations and their non-Gaussianity generated during inflation.
It was shown in Ref.~\cite{Caravano:2022epk} that the lattice calculation reproduced the curvature power spectrum in the weak backreaction regime, where the gauge-field production does not affect the background inflationary dynamics.
Furthermore, they found that non-Gaussianity is strongly suppressed in the strong backreaction regime.
However, to estimate the mass and abundance of PBHs in axion $U(1)$ inflation, one needs to perform simulations until the end of inflation, which is challenging because the strong backreaction period lasts for many e-folds.

In this paper, we consider the bumpy axion inflation model~\cite{Ozsoy:2020kat}, where the axion, identified with the inflaton, has a potential consisting of a polynomial term coming from shift-symmetry breaking in addition to a cosine term from non-perturbative effects.
In this model, the modulation caused by the non-perturbative effects makes plateaus and cliffs in the inflaton potential, which leads to a relatively short period of the strong backreaction regime. 
In previous work~\cite{Ozsoy:2020kat}, the PBH formation in this model was studied using a semi-analytical estimation of the curvature power spectrum.
Here, we investigate the curvature perturbations produced in the bumpy axion inflation model using lattice simulations and examine whether they become large enough to form PBHs.
We find that the bumpy axion inflation model generates the curvature power spectrum with a sharp peak and produces PBHs that account for the dark matter of the universe.
Moreover, we also evaluate the gravitational waves induced by the enhanced curvature perturbations. 
We find that the gravitational wave spectrum has a peak at the frequency corresponding to the peak of the curvature spectrum and that the produced gravitational waves can be probed by future experiments.

The rest of this paper is organized as follows. 
In Sec.~\ref{sec: axioninf}, we briefly review axion $U(1)$ inflation and introduce the specific models that we study.
In Sec.~\ref{sec: lattice}, we describe the numerical setup of our lattice simulations.
In Sec.~\ref{sec: result}, we show the curvature perturbations obtained by the lattice simulations for each model.
In Sec.~\ref{sec: pbh}, we evaluate the PBH abundance and the spectrum of gravitational waves induced by the enhanced curvature perturbations in bumpy axion inflation.
Finally, we conclude in Sec.~\ref{sec: conclusion}.

\section{Axion \texorpdfstring{$U(1)$}{} inflation}
\label{sec: axioninf}

First, we review the dynamics of axion $U(1)$ inflation, where a pseudo-Nambu-Goldstone boson plays the role of an inflaton.
We assume that the inflaton $\phi$ interacts with a $U(1)$ gauge field $A_{\mu}$ via the Chern-Simons interaction. 
In this case, the perturbations of the gauge field are exponentially enhanced via the Chern-Simons interaction. 
The produced gauge particles then backreact on the motion of the inflaton, increasing the power spectrum of inflaton fluctuations. 

In the following, we derive the equations of motion for the inflaton and gauge field and discuss the exponential production of gauge particles. 
The action is written as 
\begin{equation}
    S = \int \mathrm{d}^4 x \sqrt{-g}
    \left[
        \frac{M^2_\mathrm{pl}}{2}R
        -\frac{1}{2}g^{\mu\nu}(\partial_{\mu}\phi)(\partial_{\nu}\phi)
        -V(\phi)
        -\frac{1}{4}F^{\mu\nu}F_{\mu\nu}
        -\frac{\alpha}{4f}\phi F_{\mu\nu} \tilde{F}^{\mu\nu}
    \right],
    \label{eq: action}
\end{equation}
where $g\equiv\mathrm{det}(g_{\mu\nu})$ is the determinant of the spacetime metric $g_{\mu\nu}$, $M_\mathrm{pl} \simeq 2.435 \times 10^{18}$\,GeV is the reduced Planck mass, $R$ is the Ricci scalar, $V(\phi)$ is an inflaton potential, and $F_{\mu\nu} \equiv \partial_{\mu}A_{\nu}-\partial_{\nu}A_{\mu}$ is the field strength tensor of the $U(1)$ gauge field. 
In the interaction term, $\alpha$ is a dimensionless parameter, $f$ is the decay constant, $\tilde{F}^{\mu\nu} \equiv \epsilon^{\mu\nu\alpha\beta}F_{\alpha\beta}/2$ is the dual tensor of $F_{\mu\nu}$, and $\epsilon^{\mu\nu\alpha\beta}$ is the Levi-Civita tensor with $\epsilon^{0123}=1/\sqrt{-g}$.
As for the metric, we take the spatially flat Friedmann-Lema\^{i}tre-Robertson-Walker (FLRW) metric given by
\begin{equation}
    \mathrm{d}s^2
    =
    a^2(\tau)[
        -\mathrm{d}\tau^2 + \mathrm{d}x^2
        +\mathrm{d}y^2 + \mathrm{d}z^2
    ],
    \label{eq: metric}
\end{equation}
where $a$ is the scale factor, and $\tau$ is the conformal time related to the cosmic time $t$ by $a \mathrm{d} \tau = \mathrm{d} t$.
In the following, we denote derivatives with respect to $t$ and $\tau$ by $\dot{F} \equiv \partial_t F$ and $F^{\prime} \equiv \partial_{\tau} F$, respectively, with $F$ being a function of $t$ or $\tau$.

In the linear cosmological perturbation theory, we separate the inflaton field into a homogeneous part and its fluctuations as $\phi(\tau,\bm{x})=\bar{\phi}(\tau)+\delta\phi(\tau,\bm{x})$.
From Eq.~\eqref{eq: action}, the equation of motion for the homogeneous part of the inflaton is written as
\begin{equation}
    \bar{\phi}^{\prime\prime} + 2\mathcal{H}\bar{\phi}^{\prime} + a^2\frac{\mathrm{d}V}{\mathrm{d}\phi}
    =
    -a^2\frac{\alpha}{4f}\braket{F_{\mu\nu} \tilde{F}^{\mu\nu}},
    \label{eq: inflaton background eom}
\end{equation}
where $\mathcal{H} \equiv a'/a$ is the conformal Hubble parameter, and $\braket{F_{\mu\nu} \tilde{F}^{\mu\nu}}$ represents the homogeneous part of $F_{\mu\nu} \tilde{F}^{\mu\nu}$. 
On the other hand, the homogeneous part of the gauge fields vanishes as the solution of the background equation of motion.
Thus, we simply denote the gauge field by $A_\mu(\tau,\bm{x})$.
The Friedmann equation is given by
\begin{equation}
    H^2
    =
    \frac{1}{3M_\mathrm{pl}^2}(\bar{\rho}_{\phi}+\bar{\rho}_A),
    \label{eq: friedmann}
\end{equation}
where $H \equiv \dot{a}/a$ is the Hubble parameter, and $\bar{\rho}_{\phi}$ and $\bar{\rho}_A$ are the homogeneous parts of the energy densities of the inflaton and the gauge fields, respectively.
$\bar{\rho}_{\phi}$ and $\bar{\rho}_A$ are written as
\begin{align}
    \bar{\rho}_{\phi}&=\frac{1}{2}\dot{\bar{\phi}}^2+V(\bar{\phi}),
    \label{eq: inflaton background energy}
    \\
    \bar{\rho}_A&=\frac{1}{2a^4}\Braket{\sum_i(F_{0i})^2+\frac{1}{2}\sum_{i,j}(F_{ij})^2},
    \label{eq: gauge energy}
\end{align}
where $i$ and $j$ are the spatial indices, i.e., $i,j = 1,2,3$.
The equations of motion for the fluctuations of the inflaton and the gauge fields are given by
\begin{align}
    \delta\phi'' + 2\mathcal{H}\delta\phi' - \nabla^2 \delta\phi + a^2\frac{\mathrm{d}^2V}{\mathrm{d}\phi^2}\delta\phi
    &=
    -a^2\frac{\alpha}{4f}\left[
        F_{\mu\nu}\tilde{F}^{\mu\nu} -\braket{F_{\mu\nu} \tilde{F}^{\mu\nu}}
    \right],
    \label{eq: inflaton fluctuation eom}
    \\
    \bm{A}'' - \nabla^2\bm{A} - \frac{\alpha}{f}\bar{\phi}'(\nabla\times\bm{A})
    &=
    \bm{0}.
    \label{eq: gauge eom}
\end{align}
Here we take the Coulomb gauge $\nabla\cdot\bm{A}=0$ and set $A_0=0$.

To see the exponential production of the gauge field, we consider the gauge field as a quantum field and expand it as
\begin{equation}
    \hat{\bm{A}}(\tau,\bm{x})
    =
    \sum_{\lambda=\pm}\int\frac{\mathrm{d}^3k}{(2\pi)^3}
    \left[
        \hat{b}_{\bm{k},\lambda}\bm{\epsilon}_{\lambda}(\bm{k})
        A_{\lambda}(\tau,k)e^{-i\bm{k}\cdot\bm{x}}
        +\hat{b}_{\bm{k},\lambda}^\dagger\bm{\epsilon}_{\lambda}^*(\bm{k})
        A_{\lambda}^*(\tau,k)e^{i\bm{k}\cdot\bm{x}}
    \right],
    \label{eq: gauge expansion}
\end{equation}
where $\lambda =\pm$ denotes two circular polarizations of the transverse gauge field, $\hat{b}_{\bm{k},\lambda}$ ($\hat{b}_{\bm{k},\lambda}^\dagger$) is the annihilation (creation) operator, $\bm{\epsilon}_{\lambda}(\bm{k})$ is the circular polarization vector, and $A_{\lambda}$ is the gauge polarization mode function. 
The annihilation and creation operators satisfy the following commutation relations:
\begin{equation}
    [\hat{b}_{\bm{k}_1,\lambda_1},\hat{b}^{\dagger}_{\bm{k}_2,\lambda_2}]=(2\pi)^3\delta_{\lambda_1\lambda_2}\delta_\mathrm{D}(\bm{k}_1-\bm{k}_2),
    \quad 
    [\hat{b}_{\bm{k}_1,\lambda_1},\hat{b}_{\bm{k}_2,\lambda_2}]
    =[\hat{b}^{\dagger}_{\bm{k}_1,\lambda_1},\hat{b}^{\dagger}_{\bm{k}_2,\lambda_2}]
    =0,
    \label{eq: gauge commutation relation}
\end{equation}
where $\delta_{\lambda \sigma}$ is the Kronecker delta, and $\delta_\mathrm{D}$ is the three-dimensional delta function.
The polarization vectors satisfy
\begin{align}
    \begin{aligned}
        \bm{k}\cdot\bm{\epsilon}_{\pm}(\bm{k})=0
        ,\quad&
        \bm{\epsilon}^*_{\lambda_1}(\bm{k})\cdot\bm{\epsilon}_{\lambda_2}(\bm{k})=\delta_{\lambda_1\lambda_2},
        \\
        \bm{k}\times\bm{\epsilon}_{\pm}(\bm{k})&=\mp ik\bm{\epsilon}_{\pm}(\bm{k}).
    \end{aligned}
    \label{eq: polarization vector relation}
\end{align}
From Eqs.~\eqref{eq: gauge eom} and \eqref{eq: gauge expansion}, we obtain 
\begin{equation}
    A_{\pm}^{\prime\prime}(\tau,k)+\left[k^2\pm\frac{2k\xi}{-\tau}\right]A_{\pm}(\tau,k)=0
    .
    \label{eq: gauge mode eom}
\end{equation}
Here, $\xi$ is the so-called effective coupling defined by
\begin{equation}
    \xi
    \equiv 
    -a\tau\frac{\alpha\dot{\bar{\phi}}}{2f}
    \simeq
    \frac{\alpha\dot{\bar{\phi}}}{2fH},
    \label{eq: effective coupling}
\end{equation}
where we have used the relation of quasi de Sitter space $a \simeq -\frac{1}{H\tau}$.
We assume $\dot{\bar{\phi}}>0$, then $\xi>0$. 
In this case, $A_-$ experiences a tachyonic instability for $\frac{k}{aH}<2\xi$, which causes the exponential enhancement of $A_{-}$.

\subsection{Simple axion \texorpdfstring{$U(1)$}{} inflation}
\label{subsec: simple axion inflation}

In this section, we consider the simple axion $U(1)$ inflation model.
Here, we assume that inflation is a slow-roll type without specifying a concrete potential.
Following Ref.~\cite{Barnaby:2011vw}, we assume that $\xi$ is a constant, and derive the solutions for the gauge polarization mode functions $A_{\pm}$ and the curvature power spectrum. 
From Eq.~\eqref{eq: effective coupling}, $|\dot{\xi}/\xi | \simeq |\ddot{\phi}/\dot{\phi}-\dot{H}/H|$, which is much smaller than unity for slow-roll inflation.
Thus, the assumption of $\xi=\mathrm{const.}$ is valid as long as the inflaton motion is sufficiently slow.

Then, the solutions of Eq.~\eqref{eq: gauge mode eom} are written as~\cite{Anber:2009ua}
\begin{align}
\begin{aligned}
    A_+(\tau,k)&=\frac{1}{\sqrt{2k}}\left[G_0(-\xi,-k\tau)+iF_0(-\xi,-k\tau)\right],\\
    A_-(\tau,k)&=\frac{1}{\sqrt{2k}}\left[G_0(\xi,-k\tau)+iF_0(\xi,-k\tau)\right],
\end{aligned}
\label{eq: gauge mode solution (cwf)}
\end{align}
where $G$ and $F$ are the Coulomb wave functions, and the coefficients are determined so that the mode functions follow the Bunch-Davies initial condition:
\begin{equation}
    \lim_{k\tau\to-\infty}A_{\pm}(\tau,k)=\frac{1}{\sqrt{2k}}e^{-ik\tau}.
    \label{eq: gauge mode solution initial condition}
\end{equation}
Using Eq.~\eqref{eq: gauge expansion}, $\braket{F_{\mu\nu}\tilde{F}^{\mu\nu}}$ and $\bar{\rho}_A$ are written as
\begin{align}
    \braket{F_{\mu\nu}\tilde{F}^{\mu\nu}}
    &\simeq
    \frac{1}{\pi^2a^4}\int^{\infty}_0 \mathrm{d}k \,
    k^3\frac{\mathrm{d}}{\mathrm{d}\tau}|A_-(\tau,k)|^2,
    \label{eq: homogeneous backreaction 1}
    \\
    \bar{\rho}_A
    &\simeq
    \frac{1}{4\pi^2a^4}\int^{\infty}_0\mathrm{d}k \, 
    k^2\left[|A^{\prime}_-(\tau,k)|^2+k^2|A_-(\tau,k)|^2\right],
    \label{eq: gauge energy 1}
\end{align}
where we imposed the Coulomb gauge and ignored the contribution from the stable gauge polarization mode $A_+$. 
The approximated expressions of the unstable gauge polarization mode $A_-$ are given by~\cite{Barnaby:2011vw}
\begin{align}
\begin{aligned}
    A_-(\tau,k)&\simeq\frac{1}{\sqrt{2k}}\left(\frac{-k\tau}{2\xi}\right)^{\frac{1}{4}}\exp{\left(\pi\xi-2\sqrt{-2\xi k\tau}\right)},
    \\
    A_-^{\prime}(\tau,k)&\simeq\sqrt{\frac{2k\xi}{-\tau}}A_-(\tau,k).
\end{aligned}
\label{eq: cwf approximation}
\end{align}
These expressions are valid for $\frac{1}{8\xi}\lesssim-k\tau\lesssim2\xi$~\cite{Barnaby:2010vf}. 
With use of Eq.~\eqref{eq: cwf approximation}, Eqs.~\eqref{eq: homogeneous backreaction 1} and~\eqref{eq: gauge energy 1} are written as~\cite{Anber:2009ua}
\begin{align}
    \frac{\alpha}{4f}\braket{F_{\mu\nu}\tilde{F}^{\mu\nu}}&\simeq2.4\times10^{-4}\frac{\alpha}{f}\frac{H^4}{\xi^4}e^{2\pi\xi},
    \label{eq: homogeneous backreaction 2}
    \\
    \bar{\rho}_A&\simeq1.4\times10^{-4}\frac{H^4}{\xi^3}e^{2\pi\xi}.
    \label{eq: gauge energy 2}
\end{align}

When the gauge particles are exponentially produced, they affect the evolutions of $\phi$ and $a$ as backreaction.
Including the backreaction, the evolutions of $\bar{\phi}$ and $a$  are described by Eqs.~\eqref{eq: inflaton background eom} and \eqref{eq: friedmann}.
From Eqs.~\eqref{eq: homogeneous backreaction 2} and \eqref{eq: gauge energy 2}, the conditions that the backreaction of the gauge field is weak and can be ignored are given by~\cite{Barnaby:2011vw}
\begin{align}
    \frac{H^2}{2\pi|\dot{\bar{\phi}}|}
    &\ll
    13 \xi^{\frac{3}{2}}e^{-\pi\xi}
    ,
    \label{eq: backreaction condition: backreaction}
    \\
    \frac{H}{M_\mathrm{pl}}
    &\ll
    146 \xi^{\frac{3}{2}} e^{-\pi\xi}
    .
    \label{eq: backreaction condition: energy}
\end{align}
The first condition comes from the requirement that the backreaction on $\bar{\phi}$ [Eq.~\eqref{eq: homogeneous backreaction 2}]
is much smaller than the Hubble friction $3H\dot{\bar{\phi}}$, and the second condition comes from requiring that the energy density of the gauge field [Eq.~\eqref{eq: gauge energy 2}] is much smaller than the total energy density.
When both Eqs.~\eqref{eq: backreaction condition: backreaction} and~\eqref{eq: backreaction condition: energy} are satisfied, the evolutions of $\bar{\phi}$ and $a$ are not affected by the gauge field. 
This case is called the weak backreaction regime, while the case where at least one of the two conditions~\eqref{eq: backreaction condition: backreaction} and~\eqref{eq: backreaction condition: energy} is violated is called the strong backreaction regime.

Next, we discuss an increase in the curvature power spectrum caused by the backreaction of the gauge field on the inflaton fluctuations $\delta\phi$. 
The curvature perturbation $\mathcal{R}$ and the inflaton fluctuation $\delta\phi$ are related by
\begin{equation}
    \mathcal{R}=\frac{H}{\dot{\bar{\phi}}}\delta\phi.
    \label{eq: curvature perturbation}
\end{equation}
Here, $\delta\phi$ is obtained by solving Eq.~\eqref{eq: inflaton fluctuation eom} using the Mukhanov-Sasaki variable $\mathcal{F} \equiv a \delta\phi$.
Similarly to the gauge field, we quantize $\mathcal{F}$ in the momentum space by expanding it as
\begin{equation}
    \hat{\mathcal{F}}(\tau,\bm{x})
    =
    \int\frac{\mathrm{d}^3k}{(2\pi)^3}
    \left[
        \hat{a}_{\bm{k}} \mathcal{F}(\tau,\bm{k})
        e^{-i\bm{k}\cdot\bm{x}}
        +
        \hat{a}_{\bm{k}}^\dagger \mathcal{F}^*(\tau,\bm{k})
        e^{i\bm{k}\cdot\bm{x}}
    \right],
    \label{eq: F Fourier}
\end{equation}
where $\hat{a}_{\bm{k}}$ and $\hat{a}_{\bm{k}}^\dagger$ are the annihilation and creation operators satisfying
\begin{equation}
    [\hat{a}_{\bm{k}_1},\hat{a}^{\dagger}_{\bm{k}_2}]
    =
    (2\pi)^3 \delta_\mathrm{D}(\bm{k}_1-\bm{k}_2)
    , \quad 
    [\hat{a}_{\bm{k}_1},\hat{a}_{\bm{k}_2}]
    =
    [\hat{a}^{\dagger}_{\bm{k}_1},\hat{a}^{\dagger}_{\bm{k}_2}]
    =
    0.
    \label{eq: scalar commutation relation}
\end{equation}
Then, from Eq.~\eqref{eq: inflaton fluctuation eom}, the mode function $\mathcal{F}(\tau, \bm{k})$ satisfies
\begin{equation}
    \mathcal{F}''(\tau,\bm{k}) 
    +
    \left(k^2 -\frac{a''}{a} + a^2\frac{\mathrm{d}^2V}{\mathrm{d}\phi^2}\right)\mathcal{F}(\tau,\bm{k}) 
    =
    - a^3
    \frac{\alpha}{4f}(F_{\mu\nu}\tilde{F}^{\mu\nu})(\tau,\bm{k}),
\end{equation}
where $(F_{\mu\nu}\tilde{F}^{\mu\nu})(\tau,\bm{k})$ is the Fourier transform of $(F_{\mu\nu}\tilde{F}^{\mu\nu})(\tau,\bm{x})$.
$\mathcal{F}$ is written as the sum of the homogeneous solution $f_\mathrm{vac}$ and the particular solution $f_\mathrm{src}$ as $\mathcal{F}=f_\mathrm{vac}+f_\mathrm{src}$.
The former corresponds to vacuum fluctuations, and the latter to the contribution sourced by the gauge field.
$f_\mathrm{vac}$ is given by
\begin{equation}
    f_\mathrm{vac}(\tau,k)\equiv f_k(\tau) =\frac{\sqrt{-\pi\tau}}{2}H_{\nu}^{(1)}(-k\tau).
    \label{eq: inflaton fluctuation hankel}
\end{equation}
Here, $H_\nu^{(1)}$ is the Hankel function of the first kind with $\nu \equiv 3/2 +\epsilon+\eta/2$, where $\epsilon$ and $\eta$ are the slow-roll parameters defined by $\epsilon\equiv-\dot{H}/H^2$ and $\eta\equiv\dot{\epsilon}/(\epsilon H)$.
We normalized $f_k$ so that $f_k$ coincides with the mode function in the Bunch-Davies vacuum in the subhorizon limit, $f_k(\tau \to - \infty, k) = e^{-ik\tau}/\sqrt{2k}$ up to a complex phase.

On the other hand, 
$f_\mathrm{src}$ is written as
\begin{equation}
    f_\mathrm{src} (\tau,k)
    =
    \int^0_{-\infty} \mathrm{d}\bar{\tau}\,
    G_k(\tau,\bar{\tau})\left(-a^3\frac{\alpha}{4f}\right)
    (F_{\mu\nu} \tilde{F}^{\mu\nu})( \bar{\tau},
    \bm{k}),
\end{equation}
where $G_k$ is the Green function satisfying
\begin{equation}
    \left(
        \frac{\mathrm{d}^2}{\mathrm{d}\tau^2}
        + k^2 
        - \frac{a''}{a} 
        + a^2 \frac{\mathrm{d}^2V}{\mathrm{d}\phi^2}
    \right) G_k(\tau,\bar{\tau}) 
    =
    \delta(\tau-\bar{\tau}) ,
\end{equation}
and is given by
\begin{equation}
    G_k(\tau,\bar{\tau}) 
    =
    i \theta(\tau-\bar{\tau})
    [ f_k(\tau)f^*_k(\bar{\tau}) - f_k^*(\tau)f_k(\bar{\tau}) ]
    .
\end{equation}
Here, $\theta(\tau)$ is the Heaviside step function.
The power spectrum is obtained by calculating the two-point function as
\begin{align}
    \langle \mathcal{R}(\bm{k})\mathcal{R}^*(\bm{k}')\rangle 
    & =
    (2\pi)^3 \delta_\mathrm{D}(\bm{k}-\bm{k}')
    \left(\frac{H}{a\dot{\bar{\phi}}}\right)^2
    \left[ f_\mathrm{vac}(\bm{k})f^*_\mathrm{vac}(\bm{k}')
    + f_\mathrm{src}(\bm{k})f^*_\mathrm{src}(\bm{k}')\right]
    \\
    & \equiv (2\pi)^3 \delta_\mathrm{D}
    (\bm{k}-\bm{k}')\frac{2\pi^2}{k^3}\mathcal{P}_\mathcal{R}(k).
\end{align}
After some calculations, the dimensionless curvature power spectrum is written as~\cite{Barnaby:2011vw}
\begin{equation}
    \mathcal{P}_{\mathcal{R}}(k)
    =
    (-k\tau)^{n_\mathrm{s}-1}
    \left[
        \mathcal{P}_\mathrm{vac}
        +
        \mathcal{P}^2_\mathrm{vac}f_2(\xi)e^{4\pi\xi}
    \right]
    ,
    \label{eq: curvature spectra (slow-roll)}
\end{equation}
where $n_\mathrm{s}$ is the spectral index given by $n_\mathrm{s}=1-2\epsilon-\eta$,  $\mathcal{P}_\mathrm{vac}$ is the scale-invariant dimensionless curvature power spectrum obtained in the absence of the gauge field,
\begin{equation}
    \mathcal{P}_\mathrm{vac}
    =
    \left( \frac{H^2}{2\pi\dot{\bar{\phi}}} \right)^2,
    \label{eq: Pvac}
\end{equation}
and $f_2(\xi)$ is a dimensionless quantity that is numerically evaluated as~\cite{Barnaby:2011vw}
\begin{equation}
    f_2(\xi)
    \simeq
    \begin{cases}
        7.5 \times 10^{-5} \xi^{-6  } & \xi \gg 1
        \\
        3   \times 10^{-5} \xi^{-5.4} & 2 \leq \xi \leq 3
    \end{cases}.
    \label{eq: f2}
\end{equation}
The second term in Eq.~\eqref{eq: curvature spectra (slow-roll)} is the contribution of the produced gauge field and exponentially depends on $\xi$ as $\propto e^{4\pi\xi}$.
Thus, the amplitude of the curvature power spectrum is significantly enhanced for large $\xi$ compared to the case without the gauge field.

\subsection{Bumpy axion inflation}
\label{subsec: bumpy axion inflation}

We now consider bumpy axion inflation~\cite{Ozsoy:2020kat}. 
The inflaton potential has two terms coming from the shift symmetry breaking and the non-perturbative effect, and is given by
\begin{equation}
    V(\phi)=\frac{1}{2}m^2\phi^2+\Lambda^4\frac{\phi}{f}\sin{\left(\frac{\phi}{f}\right)},
    \label{eq: bumpy potential}
\end{equation}
where $m$ is the inflation mass, and $\Lambda$ is the scale of non-perturbative physics.
In this model, the second term of the potential causes a significant time dependence of $\xi$, which leads to temporary amplifications of the gauge field and the curvature power spectrum. 

\begin{figure}[t]
    \centering
    \includegraphics[width=.8\textwidth ]{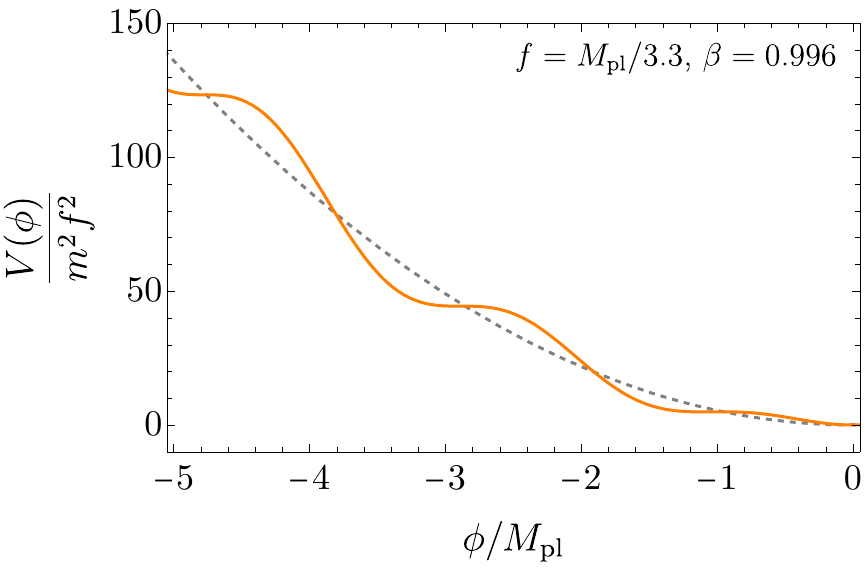}
    \caption{
        Inflaton potential $V(\phi)$ in the bumpy axion inflation model.
        The orange solid line represents Eq.~\eqref{eq: bumpy potential}, and the gray dotted line represents the mass term.
    }
    \label{fig: bumpy potential}
\end{figure}

Here, we briefly discuss the features of the potential~\eqref{eq: bumpy potential} and the inflaton motion. 
We consider the ``bumpy regime'': $\Lambda^4\lesssim m^2f^2$, where flat and steep regions alternately appear in $V(\phi)$, and the shape of the potential becomes literally ``bumpy'' compared to the case without the second term.
A typical shape of the potential is shown in Fig.~\ref{fig: bumpy potential} for  $\beta\equiv\Lambda^4/(m^2f^2)= 0.996$ and $f= M_\mathrm{pl}/3.3$. 
The inflaton slowly rolls down the flat parts of the potential, while it is accelerated and the slow-roll condition is temporarily broken in the steep parts of the potential. 
During acceleration, $\xi$ increases according to Eq.~\eqref{eq: effective coupling}, which leads to amplifications of the gauge field and the curvature power spectrum. 
However, as the inflaton quickly reaches the flat region and slows down, $\xi$ decreases. 
Therefore, the amplification is time-localized. 
Thus, in bumpy axion inflation, the time dependence of $\xi$ cannot be neglected, and hence a more sophisticated calculation is necessary.

In Ref.~\cite{Ozsoy:2020kat}, they evaluate the time evolutions of $\xi$ and the gauge field, neglecting the backreaction from the gauge field, and the semi-analytic formula for the power spectrum of curvature perturbations is presented.
However, when the curvature perturbations are large enough for PBH formation to be produced, the backreaction cannot generally be neglected.
Thus, we perform the lattice simulation to examine if the PBHs are produced in bumpy axion inflation.

\section{Lattice simulation}
\label{sec: lattice}

In this section, we introduce the lattice simulation and describe its basic methods, initial conditions, and equations of motion.
We perform three-dimensional lattice simulations by modifying the
public code \texttt{LatticeEasy}~\cite{Felder:2000hq}.
Here we use a similar setup for the lattice simulation as in Refs.~\cite{Caravano:2021bfn,Caravano:2021pgc,Caravano:2022epk}.

In the lattice simulation, we simulate the dynamics of scalar and gauge fields in a cubic box with comoving size $L$ composed of $N^3$ lattice points.
The spacing is given by $\Delta x = L/N$.
We denote the field values on the lattice points as
\begin{equation}
    f(\tau,\bm{n}),\quad n_i\in\{0,1,...,N-1\},
    \label{eq: field value on lattice}
\end{equation}
where $\bm{n}$ is the position vector of a lattice point. 
As shown in Eq.~\eqref{eq: field value on lattice}, a scalar field on the lattice has $N^3$ values, and we can track the evolution of fields by solving the equations of motion at each lattice point. 
Additionally, we impose the periodic boundary condition:
\begin{equation}
    f(\tau,\bm{n})
    =
    f(\tau,\bm{n}+N\bm{e}_x)
    =
    f(\tau,\bm{n}+N\bm{e}_y)
    =
    f(\tau,\bm{n}+N\bm{e}_z),
    \label{eq: periodic boundary condition}
\end{equation}
where $\bm{e}_x$, $\bm{e}_y$, and $\bm{e}_z$ are the unit vectors in the positive direction of each axis.

In lattice simulations, field fluctuations are automatically included in the field values on the lattice, and they need not be assumed to be small.
This allows us to follow their growth into the nonlinear regime and treat them non-perturbatively.
To extract the homogeneous quantity from a field on a lattice, $f(\tau,\bm{n})$, we calculate the lattice average as
\begin{equation}
    \bar{f}(\tau)\equiv\braket{f(\tau,\bm{n})}=\frac{1}{N^3}\sum_{\bm{n}}f(\tau,\bm{n}).
    \label{eq: lattice average}
\end{equation}
The fluctuations are then obtained by
\begin{equation}
    \delta f(\tau,\bm{n})\equiv f(\tau,\bm{n})-\bar{f}(\tau).
    \label{eq: lattice fluctuation}
\end{equation}

\subsection{Reciprocal vector and differential operators}
\label{subsec: discretization}

First, we introduce the reciprocal lattice vector
\begin{equation}
    \bm{\kappa}(\bm{m})
    =
    \frac{2\pi}{L}\bm{m},
    \quad 
    m_i 
    \in 
    \left\{-\frac{N}{2}+1,...,\frac{N}{2}\right\}.
    \label{eq: reciprocal vector}
\end{equation}
Here, $N$ is assumed to be even.
We then define the discrete Fourier transform and its inverse transform of a field $f(\tau,\bm{n})$ as
\begin{align}
    \label{eq: DFT1}
    f(\tau,\bm{\kappa})&=\sum_{\bm{n}}f(\tau,\bm{n})e^{\frac{2\pi i}{N}\bm{m}\cdot\bm{n}},
    \\
    \label{eq: DFT2}
    f(\tau,\bm{n})&=\frac{1}{N^3}\sum_{\bm{m}}f(\tau,\bm{\kappa})e^{-\frac{2\pi i}{N}\bm{m}\cdot\bm{n}}.   
\end{align}
Since we consider real fields, we assume $f(\tau,\bm{n})$ to be real, which leads to the condition on $f(\tau,\bm{\kappa})$ as
\begin{equation}
    f^*(\tau,\bm{\kappa})=f(\tau,-\bm{\kappa}),
    \label{eq: condition due to real field}
\end{equation}
where we impose the periodicity of $\bm{\kappa}(-N/2,m_y,m_z) \equiv \bm{\kappa}(N/2,m_y,m_z)$, and similarly for the $y$- and $z$-components.

Next, we define the discretized differential operator. Following Refs.~\cite{Caravano:2021pgc,Caravano:2021bfn}, we use the different definitions for single-field inflation and axion inflation. 
In the case of single-field inflation, we can eliminate gradients of the fields from the equations to solve by integrating by parts, and only the Laplacian is used to describe the evolution of the system.
Then, we use 
\begin{equation}
    \nabla^2f(\bm{n})
    =
    \frac{1}{\Delta x^2}\sum_{i=x,y,z}\sum_{c=\pm1}[f(\bm{n}+c\bm{e}_i)-f(\bm{n})].
    \label{eq: laplacian in single-field inflation}
\end{equation}
On the other hand, for axion inflation, the gradient terms are unavoidable in the equations of motion, and we need to define the gradient and Laplacian of the fields in a consistent way.
Then, we use 
\begin{align}
    \begin{aligned}
        \partial_if(\bm{n})
        &=
        \frac{f(\bm{n}+\bm{e}_i)-f(\bm{n}-\bm{e}_i)}{2\Delta x}
        ,\\
        \nabla^2f(\bm{n})
        &=
        \frac{1}{(2\Delta x)^2}\sum_{i=x,y,z}\sum_{c=\pm2}[f(\bm{n}+c\bm{e}_i)-f(\bm{n})].
    \end{aligned}
    \label{eq: laplacian in axion inflation}
\end{align}
As the discrete Fourier transform of the discretized differential operators~\eqref{eq: laplacian in single-field inflation} and~\eqref{eq: laplacian in axion inflation}, we define the effective momentum vector $\bm{k}_\mathrm{eff}$ as~\cite{Caravano:2021bfn}
\begin{equation}
    \left[k_\mathrm{eff}(\bm{m})\right]_i=\frac{2}{\Delta x}\sin{\left(\frac{\pi m_i}{N}\right)}
    \qquad (i=x,y,z),
    \label{eq: effective momentum in single-field inflation}
\end{equation}
and
\begin{equation}
    \left[k_\mathrm{eff}(\bm{m})\right]_i=\frac{1}{\Delta x}\sin{\left(\frac{2\pi m_i}{N}\right)}
    \qquad (i=x,y,z),
    \label{eq: effective momentum in axion inflation}
\end{equation}
for single-field inflation and axion inflation, respectively.
Note that $k_\mathrm{eff}$ approximately coincides with $\kappa$ for $|m_i| \ll N/2$.

\subsection{Initial condition}
\label{subsec: initial condition}

We set the initial value of the scale factor $a$ to $a_\mathrm{ini} = 1$
and that of the conformal time $\tau$ to $\tau_\mathrm{ini} = -1/H_\mathrm{ini}$ with $H_\mathrm{ini}$ being the initial value of the Hubble parameter. 
The initial value of the inflaton $\phi_\mathrm{ini}(\bm{n})$ is determined by the sum of the homogeneous component $\bar{\phi}$ and the fluctuations $\delta\phi(\bm{n})$, i.e.,
\begin{equation}
    \phi_\mathrm{ini}(\bm{n})
    =
    \bar{\phi}_\mathrm{ini} + \delta\phi_\mathrm{ini}(\bm{n})
    .
    \label{eq: initial value of inflaton field value}
\end{equation}
The initial condition for $\delta\phi(\bm{n})_\mathrm{ini}$ is taken so that $\delta\phi(\bm{n})_\mathrm{ini}$ reproduces quantum fluctuations with the analytic power spectrum given by
\begin{equation}
    P_{\delta\phi,\mathrm{ini}}(k)
    \equiv 
    \frac{2\pi^2}{k^3}\mathcal{P}_{\delta\phi,\mathrm{ini}}(k)
    =|\delta\phi_\mathrm{ini}(k)|^2
    =\frac{-\pi\tau_\mathrm{ini}}{4a_\mathrm{ini}^2}|H_{\nu}^{(1)}(-k\tau_\mathrm{ini})|^2,
    \label{eq: analytic inflaton spectrum}
\end{equation}
where $\nu$ is 
\begin{equation}
    \nu=\sqrt{\frac{9}{4}-\frac{m^2_\mathrm{eff}}{H^2_{\mathrm{ini}}}},\quad m^2_\mathrm{eff}=
    \left. 
        \frac{\mathrm{d}^2V(\phi)}{\mathrm{d}\phi^2}
    \right|_{\phi = \bar{\phi}_\mathrm{ini}}
    .
    \label{eq: hankel nu}
\end{equation}
To set the initial condition following this power spectrum, we use $X_{1,2}(\bm{m})$ and $Y_{1,2}(\bm{m})$ that are random variables uniformly distributed between 0 and 1 for each $\bm{\kappa}$, and take $\delta\phi(\bm{\kappa})_\mathrm{ini}$ as
\begin{align}
    \begin{aligned}
        \delta\phi_\mathrm{ini}(\bm{\kappa})&=\frac{1}{\sqrt{2}}|\delta\phi|_\mathrm{ini} \left[\sqrt{-\ln{X_1}}e^{i2\pi Y_1}+\sqrt{-\ln{X_2}}e^{i2\pi Y_2}\right],\\
        |\delta\phi|_\mathrm{ini} &\equiv
        \frac{L^{\frac{3}{2}}}{\Delta x^3} \frac{1}{2} \sqrt{\frac{\pi}{H_{\mathrm{ini}}}} \left|H_{\nu}^{(1)}\left(\frac{k_\mathrm{eff}}{H_{\mathrm{ini}}}\right)\right|.
    \end{aligned}
    \label{eq: initial condition of inflaton fluctuation}
\end{align}
Notice that the real and imaginary parts of the discrete Fourier mode of the inflaton fluctuations, $\mathrm{Re}[\delta\phi(\bm{\kappa})]$ and $\mathrm{Im}[\delta\phi(\bm{\kappa})]$, follow an independent Gaussian distribution $N(0, L^3 P_{\delta\phi,\mathrm{ini}}/(2 \Delta x^6))$.
Following the treatment in Ref.~\cite{Felder:2000hq}, we include two modes moving in opposite directions corresponding to $(X_1,Y_1)$ and $(X_2,Y_2)$ to realize the ``isotropic'' initial condition (see also Ref.~\cite{Figueroa:2020rrl}).
To obtain $\phi_\mathrm{ini}(\bm{n})$ via Eq.~\eqref{eq: initial value of inflaton field value}, we first substitute the values into $\delta\phi(\bm{\kappa})$ through Eq.~\eqref{eq: initial condition of inflaton fluctuation}, and then perform the inverse Fourier transform.

The initial values of the time derivative of the inflaton $\phi^{\prime}(\bm{n})$ are also determined by the sum of the homogeneous quantity $\bar{\phi^{\prime}}$ and the fluctuations $\delta\phi^{\prime}(\bm{n})$, and $\delta\phi^{\prime}(\bm{\kappa})$ are determined by the time derivative of the analytic mode function as
\begin{align}
\begin{aligned}
    \delta\phi^{\prime}_\mathrm{ini}(\bm{\kappa})
    &=
    \left[
        \left( \nu-\frac{3}{2} \right)H_{\mathrm{ini}}
        - k_\mathrm{eff}\frac{J_{\nu}J_{\nu-1}+Y_{\nu}Y_{\nu-1}}{J_{\nu}^2+Y_{\nu}^2}
    \right]
    \delta\phi_\mathrm{ini}(\bm{\kappa})\\
    &\quad
    - i\frac{\sqrt{2}}{\pi}\frac{H_\mathrm{ini} }{J_{\nu}^2+Y_{\nu}^2} |\delta\phi|_\mathrm{ini}
    \left[ 
        \sqrt{-\ln{X_1}}e^{i2\pi Y_1}
        -\sqrt{-\ln{X_2}}e^{i2\pi Y_2}
    \right],
\end{aligned}
\label{eq: initial condition of time derivative of inflaton fluctuation}
\end{align}
where $J_{\nu}$ and $Y_{\nu}$ are the Bessel function and the Neumann function, respectively, and the argument of them is $k_\mathrm{eff}/H_\mathrm{ini}$.
We use the same values of $X$ and $Y$ in Eqs.~\eqref{eq: initial condition of inflaton fluctuation} and \eqref{eq: initial condition of time derivative of inflaton fluctuation} for the same $\bm{\kappa}$.

Next, we discuss the initial condition for the gauge field.
We set $A_{0\,\mathrm{ini}}(\bm{n})=A_{0\,\mathrm{ini}}^{\prime}(\bm{n})=0$, and  furthermore the homogeneous part of the gauge field, $\bar{\bm{A}}_{\mathrm{ini}}(\bm{n})$, is assumed to be zero.
The discrete Fourier mode of the gauge field is written as
\begin{equation}
    \bm{A}_{\mathrm{ini}}(\bm{\kappa})=\bm{\epsilon}_+(\bm{\kappa})A_{+\mathrm{ini}}(\bm{\kappa})+\bm{\epsilon}_-(\bm{\kappa})A_{-\mathrm{ini}}(\bm{\kappa}).
    \label{eq: gauge polarization decomposition}
\end{equation}
Similarly to the case of the inflaton fluctuations, we use the analytic mode function of the gauge field, given by Eq.~\eqref{eq: gauge mode solution (cwf)}.
Then, the initial values of $A_\pm(\bm{\kappa})$ are given by
\begin{align}
\begin{aligned}
    A_{\pm\mathrm{ini}}(\bm{\kappa})&=\frac{1}{\sqrt{2}}|A_\pm|_{\mathrm{ini}}
    \left[\sqrt{-\ln{X_{\pm,1}}}e^{i2\pi Y_{\pm,1}}+\sqrt{-\ln{X_{\pm,2}}}e^{i2\pi Y_{\pm,2}}\right],
    \\
    |A_\pm|_\mathrm{ini}
    &=
    \frac{L^{\frac{3}{2}}}{\Delta x^3}
    \sqrt{ 
        \frac{G_0^2(\mp\xi,\frac{k_\mathbf{eff}}{H_{\mathrm{ini}}}) + F_0^2(\mp\xi,\frac{k_\mathbf{eff}}{H_{\mathrm{ini}}})}{2k_\mathrm{eff}}
    }.
\end{aligned}
\label{eq: initial condition of gauge mode}
\end{align}
We have added terms with $G_0+iF_0$ and $G_0-iF_0$ together similarly to the case of $\delta \phi_\mathrm{ini}(\bm{\kappa})$.
The initial values of the time derivative of the gauge field $\bm{A}^{\prime}_\pm(\bm{\kappa})$ are 
\begin{align}
\begin{aligned}
    A_{\pm\mathrm{ini}}^\prime(\bm{\kappa})
    &=
    \left[k_\mathrm{eff}\sqrt{1+\xi^2}\frac{G_0G_1+F_0F_1}{G_0^2+F_0^2}\pm k_\mathrm{eff}\xi-H_{\mathrm{ini}}\right]A_{\pm\mathrm{ini}}(\bm{\kappa})
    \\
    &\quad
    -\frac{i}{\sqrt{2}}
    \frac{k_\mathrm{eff}}{G_0^2+F_0^2}
    |A_\pm|_\mathrm{ini}
    \left[\sqrt{-\ln{X_{\pm,1}}}e^{i2\pi Y_{\pm,1}}-\sqrt{-\ln{X_{\pm,2}}}e^{i2\pi Y_{\pm,2}}\right],
\end{aligned}
\label{eq: initial condition for derivative of gauge field}
\end{align}
where the argument of $G_0$ and $F_0$ is $(\mp\xi,\frac{k_\mathrm{eff}}{H})$. 
Again, we use the same values of $X$ and $Y$ in Eqs.~\eqref{eq: initial condition of gauge mode} and \eqref{eq: initial condition for derivative of gauge field} for the same $\bm{\kappa}$. 
In order to get the values of $\bm{A}(\bm{n})$ and $\bm{A}^\prime(\bm{n})$, we first substitute the values into $A_\pm(\bm{\kappa})$ and $A^\prime_\pm(\bm{\kappa})$, then calculate the values of $\bm{A}(\bm{\kappa})$ and $\bm{A}^\prime(\bm{\kappa})$ through Eq.~\eqref{eq: gauge polarization decomposition}, and finally perform the inverse Fourier transform.

\subsection{Time evolution}
\label{subsec: time evolution}

In the lattice simulations, as in Refs.~\cite{Caravano:2022epk,Caravano:2024xsb}, we adopt the Lorenz gauge, $\partial^{\mu}A_{\mu}=0$, which the initial condition above satisfies. 
Then, the time evolution of the inflaton field $\phi(\bm{n})$ is described by 
\begin{equation}
    \phi^{\prime\prime}(\bm{n})+2\mathcal{H}\phi^{\prime}(\bm{n})-\nabla^2\phi(\bm{n})+a^2\frac{\mathrm{d}V}{\mathrm{d}\phi}
    =
    -\frac{1}{a^2}\frac{\alpha}{f}[\nabla\times\bm{A}(\bm{n})]\cdot[\bm{A}^{\prime}(\bm{n})-\nabla A_0(\bm{n})].
    \label{eq: inflaton eom on lattice}
\end{equation}
The equations of motion for the gauge field $A_\mu(\bm{n})$ are given by
\begin{align}
    &A_0^{\prime\prime}(\bm{n})-\nabla^2A_0(\bm{n})-\frac{\alpha}{f}[\nabla\phi(\bm{n})]\cdot[\nabla\times\bm{A}(\bm{n})]=0,
    \label{eq: gauge 0 eom on lattice}
    \\
    &\bm{A}^{\prime\prime}(\bm{n})-\nabla\bm{A}(\bm{n})+\frac{\alpha}{f}[\nabla\phi(\bm{n})]\times[\bm{A}^{\prime}(\bm{n})-\nabla A_0(\bm{n})]-\frac{\alpha}{f}\phi^{\prime}(\bm{n})[\nabla\times\bm{A}(\bm{n})]=0.
    \label{eq: gauge i eom on lattice}
\end{align}
Additionally, the scale factor $a(\tau)$ evolves as
\begin{equation}
    a^{\prime\prime}(\tau)=\frac{a^3}{6M_\mathrm{pl}^2}\braket{\rho_\phi-3p_\phi},
    \label{eq: scale factor eom on lattice}
\end{equation}
where $\rho_\phi$ and $p_\phi$ are the energy density and the pressure of the inflaton, respectively, and are given by the following equations:
\begin{align}
\begin{aligned}
    \rho_\phi(\bm{n})
    &=
    \frac{[\phi^{\prime}(\bm{n})]^2}{2a^2}-\frac{\phi(\bm{n})\nabla^2\phi(\bm{n})}{2a^2}+V(\phi(\bm{n})),
    \\
    p_\phi(\bm{n})
    &=
    \frac{[\phi^{\prime}(\bm{n})]^2}{2a^2}+\frac{\phi(\bm{n})\nabla^2\phi(\bm{n})}{6a^2}-V(\phi(\bm{n})).
    \label{eq: inflaton energy density and pressure on lattice}
\end{aligned}
\end{align}
Here, we used the relation $[\nabla\phi(\bm{n})]^2=-\phi(\bm{n})\nabla^2\phi(\bm{n})$.
We solve Eqs.~\eqref{eq: gauge 0 eom on lattice}\,--\,\eqref{eq: scale factor eom on lattice} by the fourth-order Runge-Kutta method.

We use Eq.~\eqref{eq: scale factor eom on lattice} for determining the time evolution of the scale factor instead of using the Friedmann equation given by
\begin{equation}
    \left(\frac{a'}{a}\right)^2 
    =
    \frac{\langle \rho\rangle}{3M_\mathrm{pl}^2}a^2
    ,
\end{equation}
while we check that the Friedmann equation is satisfied during the simulation. 
We also check the Lorentz gauge condition by evaluating
\begin{equation}
    \frac{\partial^{\mu}A_{\mu}}{\sqrt{\sum_\nu (\partial^{\nu}A_{\nu})^2}},
    \label{eq: lorentz gauge consistency condition}
\end{equation}
which was much smaller than unity during the simulation.

\section{Result}
\label{sec: result}

In this section, we show the results obtained from the lattice simulations. 

\subsection{Chaotic inflation}
\label{subsec: chaotic inflation}

We first present the result of the single-field slow-roll inflation. 
Since the power spectrum of the inflaton fluctuations is well-known analytically for this case, we can check whether the lattice simulation reproduces the correct power spectrum.
Here, for simplicity, we consider the chaotic inflation with only the mass term, i.e., the inflaton potential is given by
\begin{equation}
    V(\phi)
    =
    \frac{1}{2}m^2\phi^2
    .
    \label{eq: chaotic potential}
\end{equation}
We show the choice of the parameter set for the lattice simulation in Table~\ref{tab: chaotic inflation}.
This setup is almost the same as in Ref.~\cite{Caravano:2021pgc}.
\begin{table}[tbp]
    \centering
    \caption{Parameters for the lattice simulation of chaotic inflation. $\bar{\phi}_\mathrm{ini}$ and $\bar{\phi}_\mathrm{ini}^{\prime}$ are the initial values of the homogeneous component of the inflaton and its time derivative.
    }
    \label{tab: chaotic inflation}
    \vspace{3mm}
    \begin{tabular}{c| c| c| c| c|c}
        $N$ & $L$ & $\bm{k}_\mathrm{eff}$ &  $\bar{\phi}_\mathrm{ini}/M_\mathrm{pl}$ & $\bar{\phi}^{\prime}_\mathrm{ini}/M_\mathrm{pl}$ 
        & $m/M_\mathrm{pl}$
        \\[0.2em]
        \hline
        $256$& $1.4/m$ & Eq.~\eqref{eq: effective momentum in single-field inflation} & $14.5$ & $-0.815m$
        & $5.1\times10^{-6}$
    \end{tabular}
\end{table}

\begin{figure}[t]
    \centering
    \includegraphics[width=.8\textwidth ]{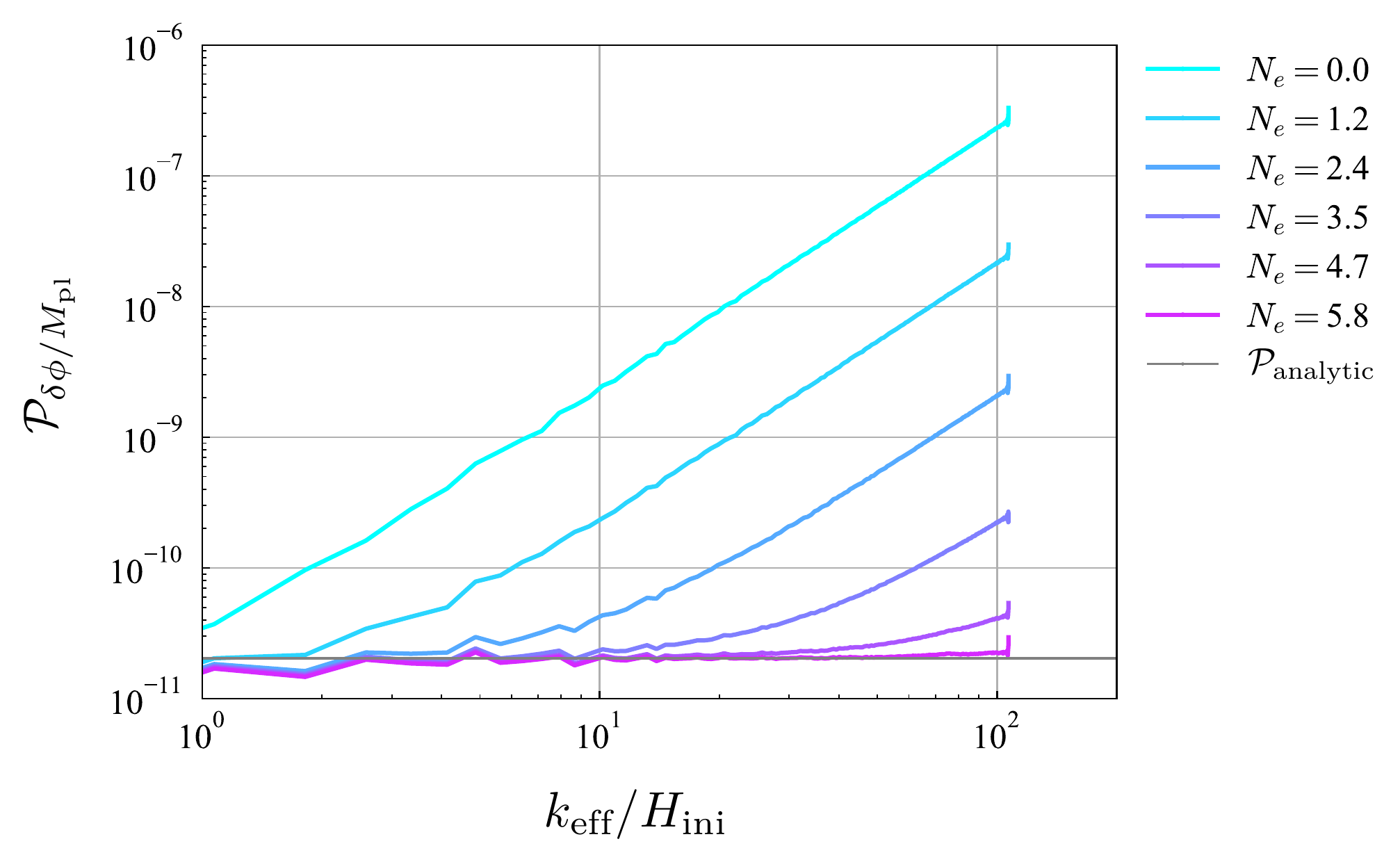}
    \caption{
        Evolution of the dimensionless power spectrum of the inflaton $\mathcal{P}_{\delta \phi/M_\mathrm{pl}} \equiv \mathcal{P}_{\delta \phi}/M_\mathrm{pl}^2$ in the lattice simulation for the chaotic inflation (colored lines). The horizontal black line represents the analytical solution.
        $N_e$ is the e-fold elapsed since the start of the lattice simulation, $k_\mathrm{eff}$ is the effective momentum vector~\eqref{eq: effective momentum in single-field inflation}, and $H_\mathrm{ini}$ is the initial value of the Hubble parameter.
    }
    \label{fig: power spectrum of chaotic inflation}
\end{figure}

We present the dimensionless power spectrum of the inflaton fluctuations in Fig.~\ref{fig: power spectrum of chaotic inflation}.
Here, $k_\mathrm{eff}/H_\mathrm{ini}$ on the horizontal axis represents the inverse of the ratio of the fluctuation scale $k_\mathrm{eff}^{-1}$ to the initial Hubble horizon $H^{-1}_\mathrm{ini}$.
At the end of the simulation $N_e = 5.8$,
all scales $k_\mathrm{eff}^{-1}$ shown in the figure are outside the horizon. 
The power spectrum at the end of the simulation (magenta line) reproduces a scale-invariant power spectrum, and its amplitude matches the analytical solution given by
\begin{equation}
    \mathcal{P}_\mathrm{analytic}=\left(\frac{H}{2\pi}\right)^2.
\end{equation}
Therefore, our lattice simulation correctly reproduces the power spectrum of the inflaton fluctuations in the case of chaotic inflation.

\subsection{Simple axion \texorpdfstring{$U(1)$}{} inflation}
\label{subsec: simple axion inflation_result}

Next, we show the result of the lattice simulation for simple axion $U(1)$ inflation discussed in Sec.~\ref{subsec: simple axion inflation}. 
We adopt the following inflaton potential:
\begin{equation}
    V(\phi) 
    =
    \frac{1}{2} m^2 \phi^2.
\end{equation}
We summarize the model and lattice parameters in Table~\ref{tab: simple axion inflation}.
The model parameters that we adopt are almost the same as those in Refs.~\cite{Caravano:2021bfn,Caravano:2022epk}. 
However, we use the Coulomb wave function~\eqref{eq: gauge mode solution (cwf)} for the initial condition of $A_+$ while Refs.~\cite{Caravano:2021bfn, Caravano:2022epk} use $A_+ = e^{-ik\tau}/\sqrt{2k}$.
This difference affects the spectrum of the stable mode of the gauge field, $A_+$, but has a negligible effect on the result of the curvature power spectrum.
We define the polarization vectors for each mode using $\bm{k}_\mathrm{eff}$ insted of $\bm{k}$ in Eq.~\eqref{eq: polarization vector relation}.
\begin{table}[tbp]
    \centering
    \caption{%
        Parameters for the lattice simulation of simple axion $U(1)$ inflation.
        As mentioned in Sec.~\ref{sec: axioninf}, we set the inflaton to move in the positive direction to ensure $\xi>0$.
    }
    \vspace{3mm}
    \begin{tabular}{c |c |c| c |c |c |c}
        $N$ & $L$ & $\bm{k}_\mathrm{eff}$ & $\bar{\phi}_\mathrm{ini}/M_\mathrm{pl}$ & $\bar{\phi}^{\prime}_\mathrm{ini}/M_\mathrm{pl}$ 
        & $m/M_\mathrm{pl}$ & $\alpha M_\mathrm{pl}/f$
        \\[0.2em]
        \hline
        $256$& $2.0/m$ & Eq.~\eqref{eq: effective momentum in axion inflation} & $-14.5$ & $0.815m$
        & $5.1\times10^{-6}$ & $42$
    \end{tabular}
    \label{tab: simple axion inflation}
\end{table}

\begin{figure}[t]
    \centering
    \includegraphics[width=.8\textwidth ]{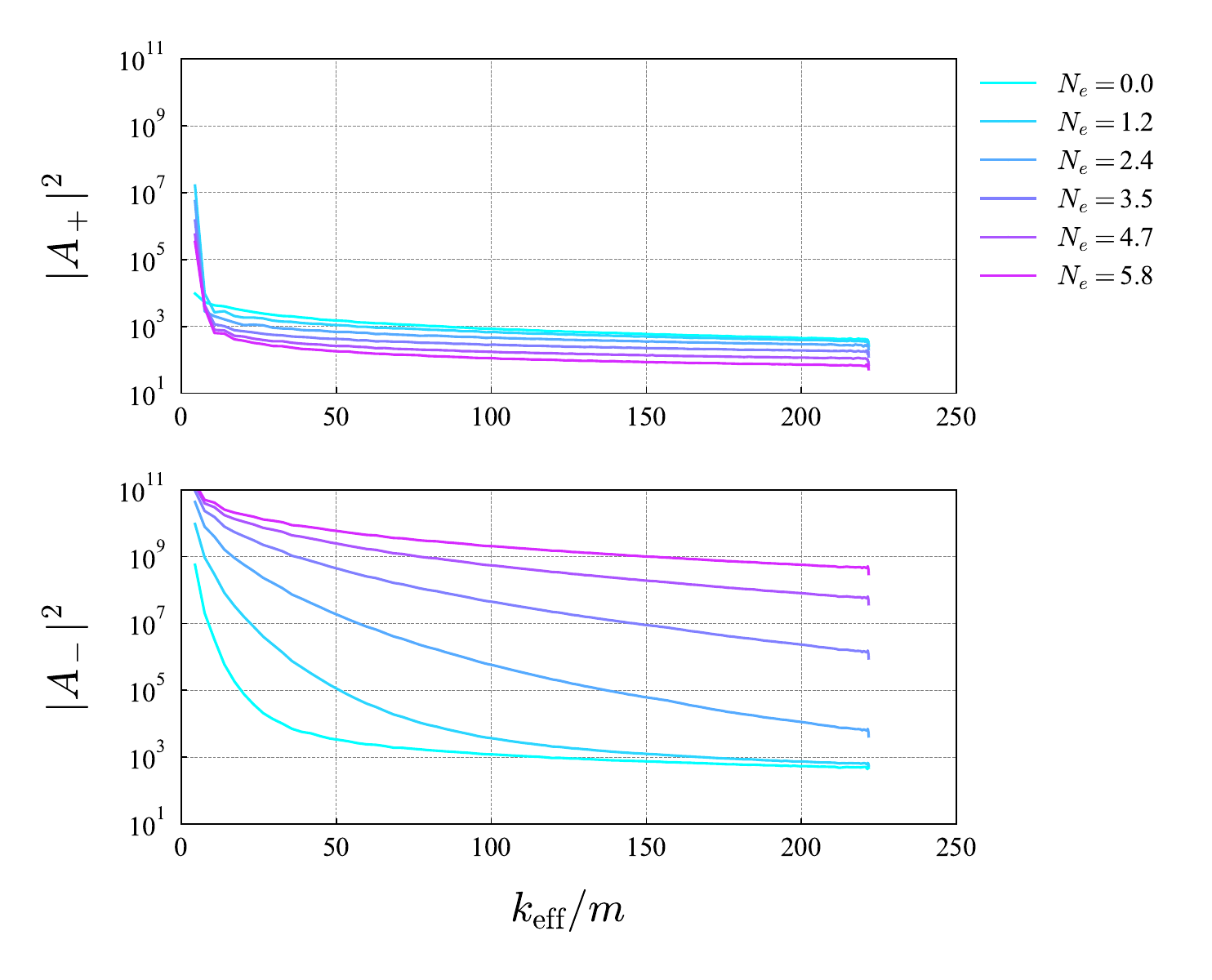}
    \caption{
        Time evolution of the spectra of the polarization mode functions of the gauge field in the lattice simulation for the simple axion $U(1)$ inflation.
        $A_+$ (top panel) and $A_-$ (bottom panel) correspond to the stable and unstable polarization modes, respectively.
    }
    \label{fig: gauge spectra}
\end{figure}

We show the spectra of the polarization mode functions of the gauge field $A_\pm$ in Fig.~\ref{fig: gauge spectra}.
Here, the unstable mode $A_-$ increases rapidly during inflation.
On the other hand, the stable mode $A_+$ slightly decreases as expected from the analytical formula~\eqref{eq: gauge mode solution (cwf)}. 
We find that $\xi$ changes from $2.88$ to $3.07$ during the simulation ($N_e =0 \,\text{--}\, 5.8$).
\begin{figure}[t]
    \centering
    \includegraphics[width=.8\textwidth ]{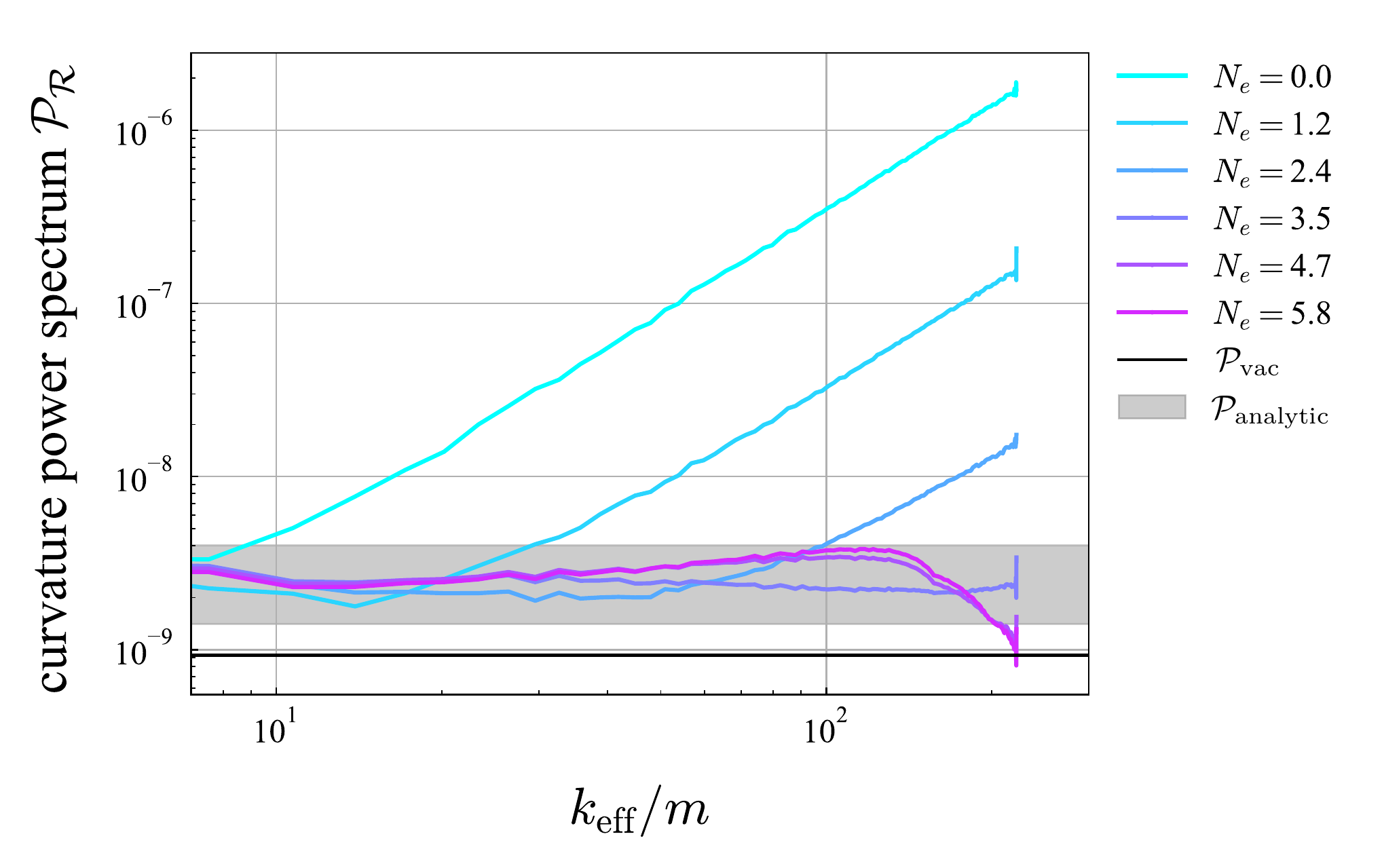}
    \caption{
        Time evolution of the dimensionless curvature power spectrum in the lattice simulation for simple axion $U(1)$ inflation (colored lines). 
        The black line represents the analytical solution for single-field inflation~\eqref{eq: Pvac}.
        The gray region represents the analytical prediction for simple axion $U(1)$ inflation~\eqref{eq: curvature spectra (slow-roll)}.
        Since this analytical solution~\eqref{eq: curvature spectra (slow-roll)} was derived under the assumption of a constant $\xi$, we show the range of the amplitude corresponding to the range of the value of $\xi$ from the start to the end of the simulation.
    }
    \label{fig: power spectrum of simple axion inflation}
\end{figure}

\begin{figure}[t]
    \centering
    \includegraphics[width=.8\textwidth ]{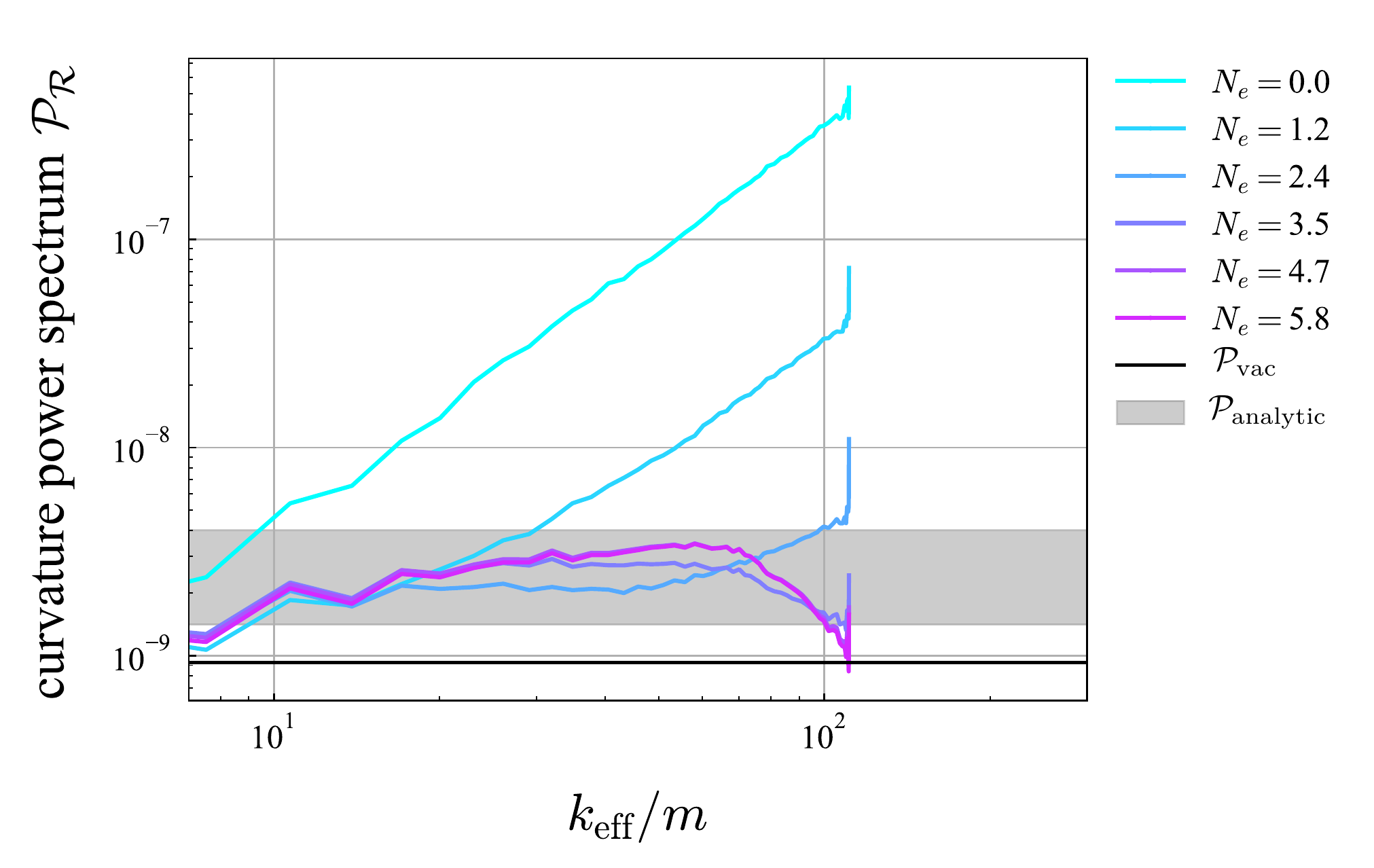}
    \caption{
        Time evolution of the dimensionless curvature power spectrum in the lattice simulation for simple axion $U(1)$ inflation with a smaller lattice box, $N=128$.
    }
    \label{fig: power spectrum of simple axion inflation (N=128)}
\end{figure}

The power spectrum of the curvature perturbations is shown in Fig.~\ref{fig: power spectrum of simple axion inflation}.
The amplitude of the power spectrum on super-horizon scales at the end of the simulation (magenta line) is larger than that in the case of the single-field chaotic inflation model~\eqref{eq: chaotic potential}.
Furthermore, the spectrum is consistent with the range of the analytic estimate~\eqref{eq: curvature spectra (slow-roll)} with $\xi =2.88\,\text{--}\,3.07$ (denoted by $\mathcal{P}_\mathrm{analytic}$). 
The decrease of the spectrum at large $k_\mathrm{eff} (\,\gtrsim 140\,m)$ is due to a UV cutoff in the lattice simulation~\cite{Caravano:2022epk}, and hence is unphysical. 
To confirm this, we show the power spectrum obtained from the simulation with a smaller lattice box ($N=128$) in Fig.~\ref{fig: power spectrum of simple axion inflation (N=128)}.
It is seen that the decrease in the spectrum occurs on a smaller scale ($k_\mathrm{eff}\gtrsim 70\,m $) due to the smaller lattice box size.
Therefore, by the lattice simulation, we confirm that the gauge field production enhances the curvature power spectrum.
The result obtained here is also consistent with that in Ref.~\cite{Caravano:2022epk}.

\subsection{Bumpy axion inflation}
\label{subsec: bumpy axion inflation_result}

In the cases above, we have shown that the lattice simulation can reproduce the power spectrum of the inflation fluctuations and curvature perturbations that were previously obtained analytically and numerically.
Now, we show the result of the lattice simulation for the bumpy axion inflation model presented in Sec.~\ref{subsec: bumpy axion inflation}. 
In this case, it is crucial to consider the time evolution of $\xi$.
Moreover, the production of the gauge field is so large that we cannot neglect the backreaction on the evolution of the homogeneous inflaton field and the Hubble parameter.
Thus, the lattice simulation is needed to obtain the precise curvature power spectrum.

\begin{table}[tbp]
    \centering
    \caption{Parameters for the lattice simulation of bumpy axion inflation.
    }
    \label{tab: bumpy axion inflation}
    \vspace{3mm}
    \begin{tabular}{c |c| c| c| c |c |c |c |c}
        $N$ & $L$ & $\bm{k}_\mathrm{eff}$ & $\bar{\phi}_\mathrm{ini}/M_\mathrm{pl}$ & $\bar{\phi}^{\prime}_\mathrm{ini}/M_\mathrm{pl}$ & $m/M_\mathrm{pl}$ 
        & $f/M_\mathrm{pl} $ & $\alpha$
        & $\beta$
        \\
        \hline
        $256$& $3.5/m$ & Eq.~\eqref{eq: effective momentum in axion inflation} & $-4.427$ & $0.3618m$ & $5.1\times10^{-6}$ 
        & 1/3.3 & 6.7
        & $0.996$
    \end{tabular}
\end{table}

\begin{figure}[t]
    \centering
    \includegraphics[width=.285\textwidth]{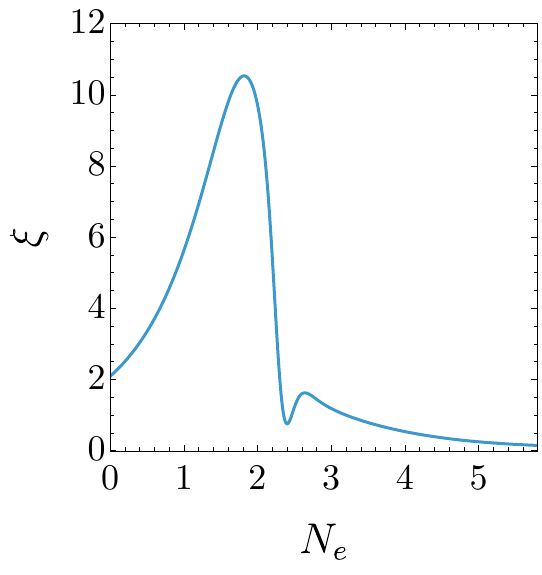}
    \includegraphics[width=.3\textwidth]{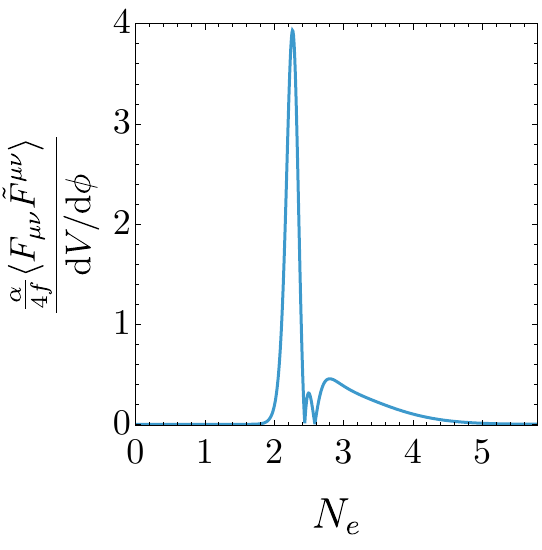}
    \includegraphics[width=.32\textwidth]{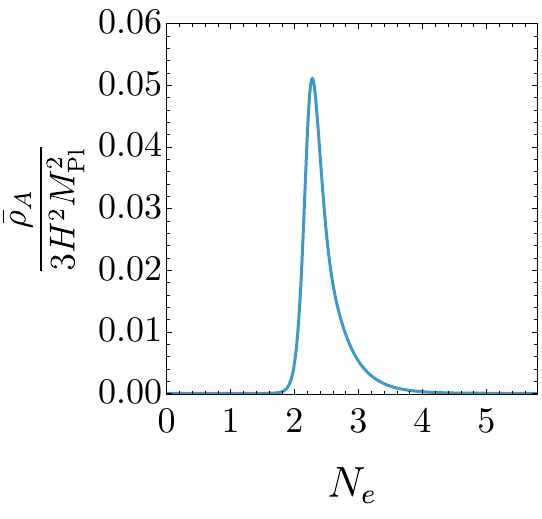}
    \caption{
        Time evolution of $\xi$ in bumpy axion inflation and parameters representing the strength of the backreaction.
    }
    \label{fig: xi & backreaction}
\end{figure}

\begin{figure}[t]
    \centering
    \includegraphics[width=.8\textwidth ]{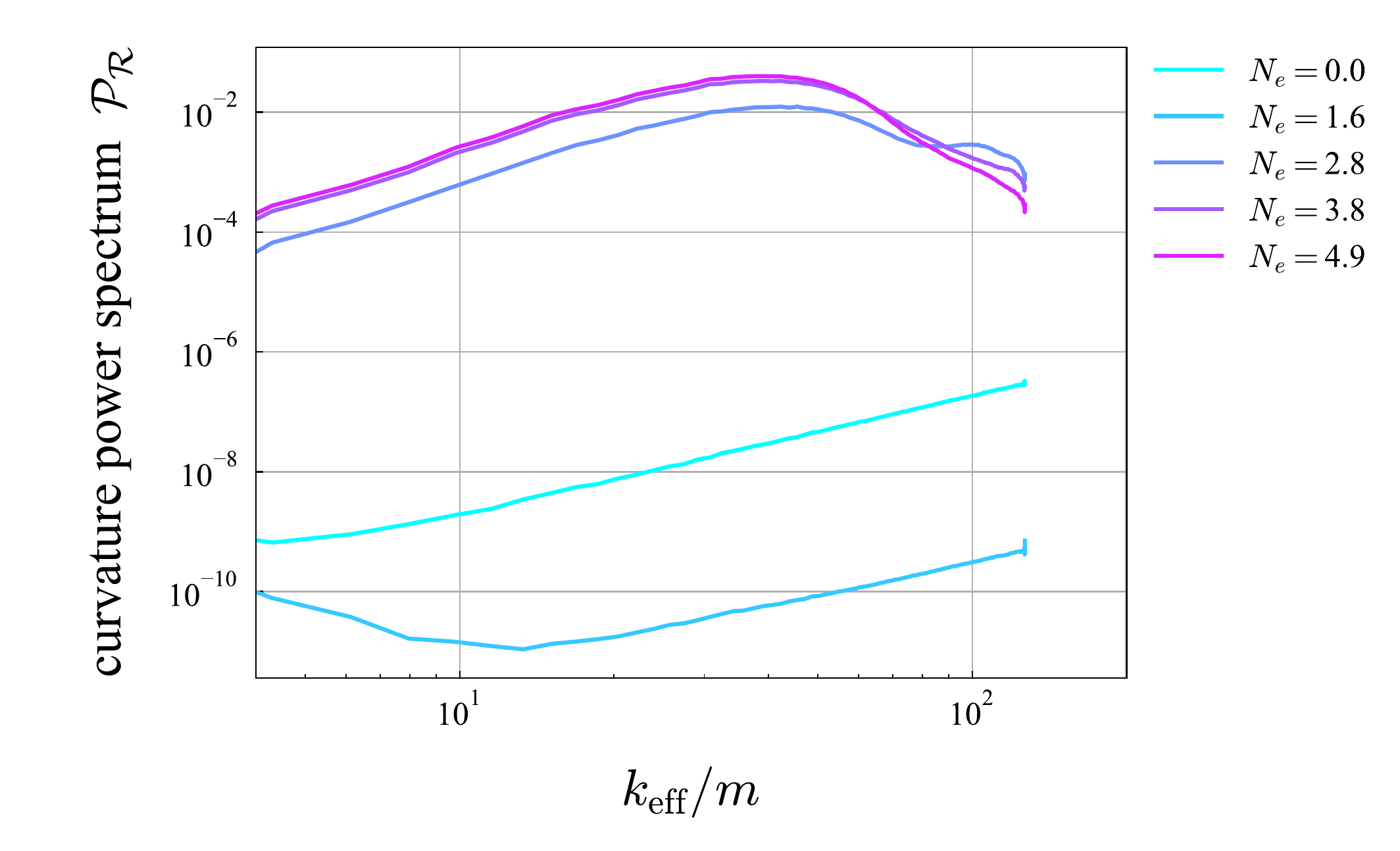}
    \caption{
        Time evolution of the dimensionless curvature power spectrum in the lattice simulation for the bumpy axion inflation.
    }
    \label{fig:power_spectrum_bumpy}
\end{figure}

We show the lattice and model parameters in Table~\ref{tab: bumpy axion inflation}.
The setup for the initial condition of $A_+$ and the polarization vector is the same as that in Sec.~\ref{subsec: simple axion inflation_result}.
In the left panel of Fig.~\ref{fig: xi & backreaction}, we show the time evolutions of $\xi$.
Here, $\xi$ increases rapidly when the inflaton rolls down the steep region of the potential and decreases after it reaches the flat region.
We also show the quantities characterizing the strength of backreaction in the center and right panels of Fig.~\ref{fig: xi & backreaction} (see also Eqs.~\eqref{eq: inflaton background eom} and \eqref{eq: friedmann}).
After $\xi$ reaches its maximum value, the backreaction of the gauge field on the inflaton becomes strong and exceeds the gradient of the inflaton potential, leading to a rapid decrease in $\xi$.
Subsequently, $\xi$ exhibits oscillatory behavior and follows the value in the weak backreaction regime.
On the other hand, the energy density of the gauge field remains subdominant even in the strong backreaction regime.
The dimensionless curvature power spectrum is shown in Fig.~\ref{fig:power_spectrum_bumpy}.

To relate the wavenumbers in the lattice simulation to the physical quantities, we need to fix the horizon scale at the initial time of the simulation.
In the following, we use 
\begin{align}
    H_\mathrm{CMB}
    =
    1.943 m
    \ , \quad 
    N_{e,\mathrm{CMB}}
    =
    \ln a_\mathrm{CMB}
    =
    -31.37
    \ ,
\end{align}
where the subscript CMB represents quantities at the horizon exit of the CMB scale, $k_\mathrm{CMB} = 0.05\,\mathrm{Mpc}^{-1}$.
With these quantities, we obtain
\begin{align}
    m 
    =
    \frac{m}{H_\mathrm{CMB}} \frac{k_\mathrm{CMB}}{a_\mathrm{CMB}}
    \simeq 
    1.06 \times 10^{10}\,\mathrm{Mpc}^{-1}
    \ .
\end{align}

\section{PBH formation and gravitational waves}
\label{sec: pbh}

Using the curvature perturbations (Fig.~\ref{fig:power_spectrum_bumpy}) obtained in the previous section, we evaluate the PBH mass function.
When the overdensity regions with scale $k$ enter the horizon, they collapse into black holes if the density contrast is larger than the critical value $\delta_c$, for which we use $\delta_c \simeq 0.41$~\cite{Harada:2013epa}.
The PBH mass is roughly given by the horizon mass $M_H$.
We represent the PBH mass as $M_\mathrm{PBH} = \gamma M_H$ with a coefficient $\gamma$, which is set to $0.2$ as a typical value~\cite{Carr:1975qj}.
The fraction $\beta$ of the PBH energy density to the total density at formation is estimated by using the Press-Schechter formula~\cite{Press:1973iz} as
\begin{equation}
    \label{eq:PBH_fraction}
    \beta 
    =
    \gamma \int_{\delta_c}^\infty \frac{\mathrm{d}\delta}{\sqrt{2\pi \sigma^2}}
    \exp\left(-\frac{\delta^2}{2\sigma^2}\right),
\end{equation}
where $\sigma^2$ is the variance of the density perturbations smoothed over the horizon scale ($k^{-1} = 1/(aH)$), which is given by
\begin{equation}
    \label{eq:variance}
    \sigma^2 = \int \frac{\mathrm{d}q}{q} \tilde{W}(k^{-1},q)\frac{16}{81}
    \left(\frac{q}{k}\right)^4\mathcal{P}_\mathcal{R} (q).
\end{equation}
Here, $\tilde{W}(R,\bm{q})$ is the Fourier transform of the window function $W(R,\bm{x})$, for which we adopt the Gaussian window function,

\begin{align}
    W(R,x) & = \frac{1}{(2\pi)^{3/2}R^3}
        \exp\left(-\frac{x^2}{2R^2}\right) ,
    \\[0.4em]
    \tilde{W}(R,q) & = \exp\left(-\frac{k^2 R^2}{2}\right).
\end{align}
The present abundance of PBHs is then written as
\begin{align}
    \label{eq:pbh_abundance}
    f_\mathrm{PBH}(M_\mathrm{PBH}) 
    &=
    \frac{1}{\rho_{\mathrm{DM},0}}
    \frac{\mathrm{d}\rho_{\mathrm{PBH},0}}{\mathrm{d}\ln M_\mathrm{PBH}}
    =
    \beta \left(\frac{T_\mathrm{form}}{T_\mathrm{eq}}\right)
    \frac{\Omega_\mathrm{M}}{\Omega_\mathrm{DM}} 
    \nonumber \\
    &= 
    \left(\frac{\beta}{5.9\times 10^{-9}}\right)
    \left(\frac{\gamma}{0.2}\right)^{1/2}
    \left(\frac{g_*}{106.75}\right)^{-1/4}
    \left(\frac{M_\mathrm{PBH}}{M_\odot}\right)^{-1/2},
\end{align}
where $\rho_{\mathrm{DM},0}$ ($\rho_{\mathrm{PBH},0}$) is the present energy density of the dark matter (PBHs), $\Omega_\mathrm{DM}$ ($\Omega_\mathrm{M}$) is the density parameter of the dark matter (total matter), $T_\mathrm{form}$ is the temperature at the PBH formation, $T_\mathrm{eq}$ is the temperature at the matter-radiation equality, and $g_*$ is the relativistic degrees of freedom for the energy density at $T_\mathrm{form}$.
Here, $\beta$ for fixed $M_\mathrm{PBH}$ is evaluated with the horizon scale $k = aH$ at $T_\mathrm{form}$, and we have used the relation between $T_\mathrm{form}$ and $M_\mathrm{PBH}$,
\begin{equation}
    M_\mathrm{PBH}
    =
    0.95\times 10^{-14}\, M_\odot
    \left(\frac{\gamma}{0.2}\right)
    \left(\frac{g_*}{106.75}\right)^{-1/2}
    \left(\frac{T_\mathrm{form}}{10^3\,\mathrm{TeV}}\right)^{-2}.
\end{equation}
In deriving these numerical relations, we used $T_\mathrm{eq} = 0.8$\,eV, $\Omega_\mathrm{M}h^2 = 0.142$, and $\Omega_\mathrm{DM}h^2 = 0.12$, $g_{*,\mathrm{eq}} = 3.36$, and $g_{*s,\mathrm{eq}} = 3.91$.
Here, $h$ is the reduced Hubble parameter, and $g_{*,\mathrm{eq}}$ and $g_{*s,\mathrm{eq}}$ are the relativistic degrees of freedom for the energy density and entropy density at $T_\mathrm{eq}$, respectively.

Before estimating the abundance of the PBHs from the result of the lattice simulation of the bumpy axion inflation model, we comment on the non-Gaussianity of the curvature perturbations.
Since the curvature perturbations are produced from non-linear fluctuations of the gauge field, one expects large non-Gaussianity, which is shown in Ref.~\cite{Caravano:2022epk} for the weak backreaction case.
However, for the strong backreaction case, it is found that the produced curvature perturbations follow Gaussian statistics due to the central limit theorem~\cite{Caravano:2022epk}.
Actually, in the lattice simulations, we confirmed that the curvature perturbation approaches a Gaussian distribution after the strong backreaction becomes effective.
We also find that the probability distribution of the inflaton fluctuation evolves in time and becomes non-Gaussian.
This behavior can be understood from the nonlinear relation between the curvature perturbation and the inflaton fluctuation, where higher-order terms become relevant for fluctuations sufficiently enhanced to produce PBHs.
Therefore, we assume that the curvature perturbations obtained in the previous section can be approximated to be Gaussian, although there could remain a small non-Gaussianity.

\begin{figure}[t]
    \centering
    \includegraphics[width=.8\textwidth]{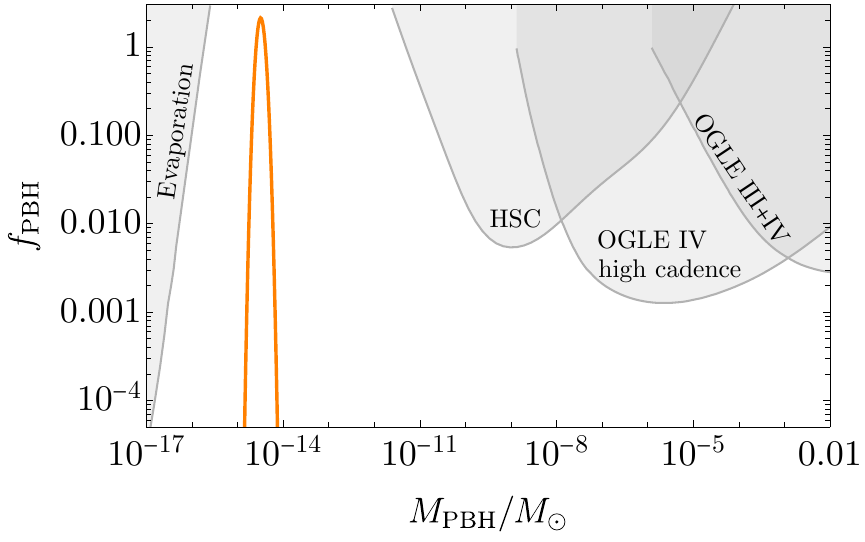}
    \caption{
        PBH abundance in the bumpy axion inflation model with the parameters shown in Table~\ref{tab: bumpy axion inflation}.
        The gray-shaded regions represent the observational upper bounds on the PBH abundance by microlensing~\cite{Croon:2020ouk,Mroz:2024mse,Mroz:2024wia} and PBH evaporation~\cite{Mittal:2021egv,Tan:2024nbx}.
        A part of the limits shown here are taken from PBHbounds~\cite{PBHbounds}.
    }
    \label{fig: fPBH}
\end{figure}

The PBH abundance evaluated for $\mathcal{P}_\mathcal{R}$ obtained in the lattice simulation of bumpy axion inflation is shown in Fig.~\ref{fig: fPBH} together with the observational constraints from microlensing~\cite{Croon:2020ouk,Mroz:2024mse,Mroz:2024wia} and PBH evaporation~\cite{Mittal:2021egv,Tan:2024nbx}.
The abundance shown here corresponds to $\rho_{\mathrm{PBH},0}/\rho_{\mathrm{DM},0} \simeq 1$.
It is seen that bumpy axion inflation can produce dark matter PBHs with masses of $\mathcal{O}(10^{-15}\,\text{--}\,10^{-14})M_\odot$ without conflicting with the observational constraints.

In bumpy axion inflation, the perturbations of the gauge field are exponentially enhanced on certain scales.
In addition to the inflaton fluctuations, they induce primordial gravitational waves as a second-order effect.
Further, primordial gravitational waves are also induced by the enhanced curvature perturbations after inflation.
Actually, it is shown that the latter gives the dominant contribution to gravitational waves in the previous study~\cite{Ozsoy:2020kat}.
Here, we evaluate the gravitational wave spectrum induced by the curvature perturbations following Ref.~\cite{Ozsoy:2020kat}.

The power spectrum of the induced tensor perturbations is given by 
\begin{align}
    \mathcal{P}_\lambda(\tau, k)
    =
    2 \int _0^\infty \mathrm{d} t
    \int_{-1}^1 \mathrm{d} s\,
    \left( \frac{t (2+t)(1-s^2)}{(1-s+t)(1+s+t)} \right)^2
    \overline{I^2(s,t,k\tau)} 
    \mathcal{P}_\mathcal{R}\left(\frac{1-s+t}{2}k\right)
    \mathcal{P}_\mathcal{R}\left(\frac{1+s+t}{2}k\right)
    \ ,
\end{align}
where $I(s,t,x)$ includes the time integral of the scalar source, and the oscillation average of $I^2$, $\overline{I^2}$, in the late time limit is given by
\begin{align}
    \overline{I^2(s,t,x \to \infty)} 
    =
    &\frac{288 [-5 + s^2 + t(2+t)]^2}{x^2 (1-s+t)^6 (1+s+t)^6}
    \left(
        \frac{\pi^2}{4}[ -5+s^2 + t(2+t)]^2
        \Theta(t-(\sqrt{3} - 1))
    \right.
    \nonumber \\        
    &\left. 
        +
        \left[ 
            - (1-s+t)(1+s+t)
            + \frac{1}{2} [-5+s^2 + t(2+t)] \ln 
            \left| 
                \frac{-2+t(2+t)}{3-s^2}
            \right|
        \right]^2
    \right)
    \ .
\end{align}
During the radiation-dominated era after the generation of gravitational waves, the fractional energy density spectrum of the gravitational waves is given by 
\begin{align}
    \Omega_\mathrm{GW,RD}(\tau, k)
    \equiv 
    \frac{1}{\rho_\mathrm{tot}} \frac{\mathrm{d} \rho_\mathrm{GW}}{\mathrm{d} \ln k}
    =
    \frac{(k\tau)^2}{12} \mathcal{P}_\lambda(\tau, k)
    \ ,
\end{align}
for subhorizon modes.
Here, $\rho_\mathrm{tot}$ is the total energy density of the universe, which is dominated by radiation.
The density parameter at the current time is given by 
\begin{align}
    \Omega_\mathrm{GW} h^2
    =
    \Omega_\mathrm{R} h^2 
    \frac{g_*}{g_{*0}} \left( \frac{g_{*s0}}{g_{*s}} \right)^{4/3}
    \Omega_\mathrm{GW,RD} h^2
    \ ,
\end{align}
where $h$ is the reduced Hubble constant, $g_{*s}$ is the relativistic degrees of freedom for the entropy density, and $\Omega_\mathrm{R}$ is the current density parameter of the radiation.

\begin{figure}[t]
    \centering
    \includegraphics[width=.8\textwidth ]{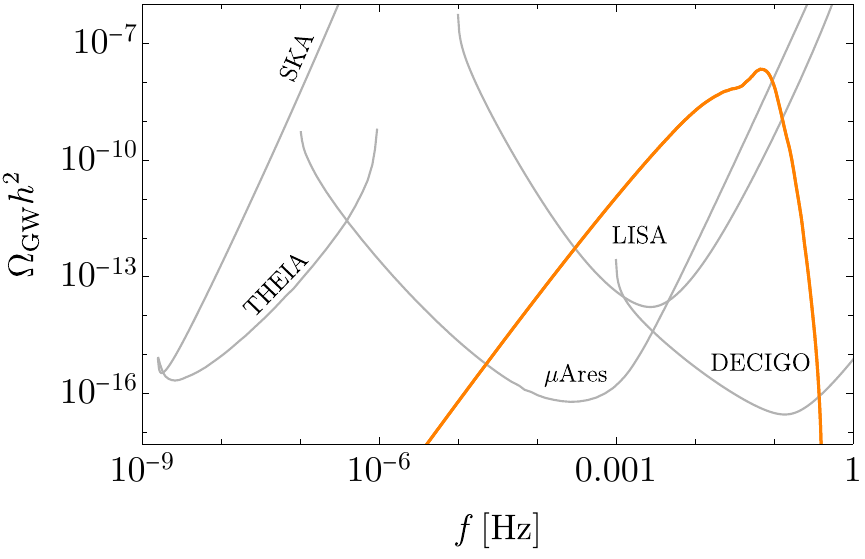}
    \caption{
        Gravitational wave spectrum in the bumpy axion inflation model.
        The gray lines represent future sensitivities of the Square Kilometre Array (SKA)~\cite{Janssen:2014dka,Weltman:2018zrl}, Telescope for Habitable Exoplanets and Interstellar/Intergalactic Astronomy (THEIA)~\cite{Garcia-Bellido:2021zgu}, $\mu$Ares~\cite{Sesana:2019vho}, Laser Interferometer Space Antenna (LISA)~\cite{LISA:2017pwj}, and DeciHertz Interferometer Gravitational-Wave Observatory (DECIGO)~\cite{Kawamura:2006up}.
        We take the sensitivities of SKA, LISA, and DECIGO from Ref.~\cite{Schmitz:2020syl}. 
    }
    \label{fig: GW}
\end{figure}

In Fig.~\ref{fig: GW}, we show $\Omega_\mathrm{GW} h^2$ evaluated using the curvature perturbations obtained in Sec.~\ref{subsec: bumpy axion inflation_result}.
Here, we used $g_{*0} = 3.36$, $g_{*s0} = 3.91$, $g_* = g_{*s} = 106.75$, and $\Omega_\mathrm{R} h^2 \simeq 4.2 \times 10^{-5}$~\cite{Planck:2018vyg}.
The produced gravitational waves are within the sensitivity range of future space-based interferometers such as $\mu$Ares~\cite{Sesana:2019vho}, LISA~\cite{LISA:2017pwj}, and DECIGO~\cite{Kawamura:2006up}.

\section{Conclusion}
\label{sec: conclusion}

We have studied PBH formation in bumpy axion inflation using lattice simulations.
In this class of models, the tachyonic production of gauge particles can strongly backreact on the inflaton dynamics.
Therefore, lattice simulations are necessary to follow the coupled evolution of the inflaton and gauge fields and to evaluate the curvature perturbations beyond the perturbative treatment.
In this work, we focused on bumpy axion inflation.
In this model, the inflaton is accelerated when it passes through the steep regions of the potential, and the effective coupling $\xi$ temporarily increases.
This leads to efficient production of gauge particles, which subsequently backreact on the background inflatonary dynamics.

As a validation of the numerical setup, we first applied our lattice simulations to chaotic inflation and simple axion $U(1)$ inflation.
We confirmed that our lattice simulations correctly reproduce the known result of the curvature power spectrum in both cases.
We then performed the lattice simulation for the bumpy axion inflation.
As a result, we found that the curvature power spectrum develops a sharp peak with a sufficiently large amplitude to lead to PBH formation.

Then, we translated the e-folds in the lattice simulation into the physical wavenumber and evaluated the resulting PBH abundance.
Consequently, we found that bumpy axion inflation can produce PBHs with an abundance sufficient to account for all dark matter in the mass region of $M_\mathrm{PBH} = \mathcal{O}(10^{-15}\,\text{--}\,10^{-14}) M_\odot$.

In axion $U(1)$ inflation, the gravitational waves are also produced from the enhanced perturbations of the gauge field and the curvature.
We evaluated the dominant contribution from the curvature perturbations and showed that the produced gravitational waves can be probed in future space-based interferometers around the $\mu$Hz or mHz range.

\begin{acknowledgments}
K.M. would like to thank Junseok Lee for helpful discussions and comments.
This work was supported by JSPS KAKENHI Grant Nos. 20H05851(M.K.), 21K03567(M.K.), 25K07297(M.K.), 23KJ0088(K.M.), and 24K17039(K.M.), and World Premier International Research Center Initiative (WPI Initiative), MEXT, Japan (M.K.).
\end{acknowledgments}

\bibliographystyle{JHEP}
\bibliography{Ref}

\end{document}